\documentclass[lettersize,journal]{IEEEtran}
\usepackage{amsmath,amsfonts}
\usepackage{algorithmic}
\usepackage{algorithm}
\usepackage{array}
\usepackage[caption=false,font=normalsize,labelfont=sf,textfont=sf]{subfig}
\usepackage{textcomp}
\usepackage{stfloats}
\usepackage{url}
\usepackage{verbatim}
\usepackage{graphicx}
\usepackage{cite}
\usepackage{caption}
\usepackage{multirow}
\usepackage{subfig}
\usepackage{caption}
\usepackage{bm}
\usepackage{makecell} 
\usepackage{epstopdf}

\begin{document}

\title{A Unified RCS Modeling of Typical Targets for 3GPP ISAC Channel Standardization and Experimental Analysis}

\author{Yuxiang Zhang, Jianhua Zhang, Xidong Hu, Jiwei Zhang, Hongbo Xing, Huiwen Gong, Shilin Luo, Yifeng Xiong, Li Yu, Zhiqing Yuan, Guangyi Liu, Tao Jiang
        
\thanks{This research is supported by Young Scientists Fund of the National Natural Science Foundation of China (62201087, 62101069), National Key R\&D Program of China (2023YFB2904803), National Natural Science Foundation of China (62341128) and Beijing University ot Posts and Telecommunications-China Mobile Research Institute Joint innovation Center.(\emph {Corresponding author: Jianhua Zhang.})}

\thanks{Yuxiang Zhang, Jianhua Zhang, Xidong Hu, Jiwei Zhang, Hongbo Xing, Huiwen Gong, Shilin Luo  and Li Yu are with the State Key Laboratory of Networking and
Switching Technology, Beijing University of Posts and Telecommunications, Beijing 100876, China (e-mail: zhangyx@bupt.edu.cn; jhzhang@bupt.edu.cn; hxd@bupt.edu.cn; rediaose@bupt.edu.cn; hbxing@bupt.edu.cn;  birdsplan@bupt.edu.cn; luoshilin@bupt.edu.cn; li.yu@bupt.edu.cn).}

\thanks{Yifeng Xiong is with the School of Information and Electronic Engineering, Beijing University of Posts and Telecommunications, Beijing 100876, China (e-mail: yifengxiong@bupt.edu.cn).}

\thanks{Zhiqiang Yuan is with the National Mobile Communications Research Laboratory, School of Information Science and Engineering, Southeast University, Nanjing 210096, China (email: zqyuan@seu.edu.cn).}

\thanks{Guangyi Liu and Tao Jiang are with the Future Research Laboratory, China Mobile Research Institute, Beijing 100053, China (e-mail: liuguangyi@chinamobile.com; jiangtao@chinamobile.com).}

}



\maketitle

\begin{abstract}

Accurate radar cross section (RCS) modeling is crucial for characterizing target scattering and improving the precision of Integrated Sensing and Communication (ISAC) channel modeling. Existing RCS models are typically designed for specific target types, leading to increased complexity and lack of generalization. This makes it difficult to standardize RCS models for 3GPP ISAC channels, which need to account for multiple typical target types simultaneously. Furthermore, 3GPP models must support both system-level and link-level simulations, requiring the integration of large-scale and small-scale scattering characteristics. To address these challenges, this paper proposes a unified RCS modeling framework that consolidates these two aspects. The model decomposes RCS into three components: (1) a large-scale power factor representing overall scattering strength, (2) a small-scale angular-dependent component describing directional scattering, and (3) a random component accounting for variations across target instances. We validate the model through mono-static RCS measurements for UAV, human, and vehicle targets across five frequency bands. The results demonstrate that the proposed model can effectively capture RCS variations for different target types. Finally, the model is incorporated into an ISAC channel simulation platform to assess the impact of target RCS characteristics on path loss, delay spread, and angular spread, providing valuable insights for future ISAC system design.
\end{abstract}


\section{Introduction}

\IEEEPARstart{I}{ntegrated} Sensing and Communication (ISAC) is a technology that integrates communication and sensing capabilities by sharing spectrum, hardware, and signal processing resources \cite{ref:Foreign_experts,REF1}. It enables high-precision localization, target detection, and environmental sensing \cite{REF5, ref:fanTIT}. Recognized in IMT-2030 as one of the six key application scenarios for 6G \cite{R1_IMT}, ISAC is set to become a critical enabler of future 6G systems \cite{REF2}.

The successful deployment of the ISAC system relies on an accurate and effective channel model \cite{ref:jialincommag,ref:3DMIMOJSAC}.
Unlike traditional communication channels, the ISAC places emphasis on the channel associated with the sensing targets (STs), where the accurate modeling of STs is critical \cite{REF2_1_1,ref:liufanJSAC}. 
The radar cross section (RCS) is defined as the effective area of a target that captures and reflects the electromagnetic wave energy \cite{REF_2_4,ref:wenjunletter}, thus characterizing the scattering characteristics of the target. As a crucial component in target modeling, RCS directly affects key tasks such as detection, localization, and tracking. 
RCS is highly sensitive to factors such as the target's geometry, material properties, surface roughness, orientation, frequency, and polarization, resulting in significant differences between various types of targets. This is particularly important for key ISAC targets, such as unmanned aerial vehicles (UAVs), vehicles, and humans, etc, which are defined as critical targets by the Third Generation Partnership Project (3GPP) \cite{REF_4}. 

3GPP established a work item proposal on ISAC in 2021, officially initiating research and standardization efforts \cite{REF_4}. This working group outlined several requirements for ISAC channel modeling . First, the channel model must account for both large-scale and small-scale effects \cite{3GPPPPPP}. Large-scale models primarily focus on signal attenuation and shadowing effects over long distances, while small-scale models describe phenomena such as fast fading and multipath effects over short distances \cite{ref:yamengISACmodel,ref:yamengshare}. Therefore, in the ISAC channel model, the RCS of the target must be modeled in both large-scale and small-scale models. Thus, understanding the RCS characteristics of different targets is essential for developing accurate and scalable ISAC channel models.

To address the challenges of RCS modeling, extensive theoretical and experimental research has been conducted.   Computational electromagnetics (CEM) methods, including finite difference time domain (FDTD) \cite{REF_dianci1}, the moment method (MoM) \cite{REF_dianci2}, and the finite element method (FEM) \cite{REF_dianci3}, solve Maxwell’s equations to emulate electromagnetic interactions with targets, offering high accuracy in RCS calculations.   However, these methods face significant computational challenges for large targets having complicated shapes.   High-frequency methods such as physical optics (PO) \cite{REF_dianci4}, geometric optics (GO) \cite{REF_dianci5}, and the physical theory of diffraction (PTD) \cite{REF_dianci6} provide approximate solutions with reduced computational complexity but sacrifice accuracy and require corrections for specific scenarios.   Experimental measurements, on the other hand, provide more realistic insights into the characteristics of scattering.   For example, RCS measurements in anechoic chambers reveal that UAVs exhibit significant angular variations despite small overall RCS values \cite{REF12, REF12_again, REF13}, while human targets show minimal angular dependence under monostatic conditions, reflecting omnidirectional scattering \cite{REF15, REF16, REF17}.   Vehicle measurements, on the contrary, reveal a strong angular dependence, with peak RCS values concentrated on flat surfaces such as the front, rear and sides \cite{REF15, REF18, REF19}. These theoretical and experimental studies form the foundation for understanding ISAC target scattering properties.


It can be seen that existing RCS modeling methods are typically designed for specific target types, and there is currently no unified framework that accommodates multiple target types. Modeling RCS characteristics for different targets separately increases the complexity of target channel models and hinders the standardization of ISAC channels. Specifically, building separate models for different targets complicates the modeling process and makes it difficult to handle the RCS characteristics of multiple target types in a unified manner. Additionally, for system-level or link-level ISAC performance evaluation, RCS models must account for both large-scale and small-scale scattering characteristics. However, due to the many factors influencing RCS, accurately parameterizing both scales is challenging. Finally, while electromagnetic simulations can compute RCS, the computational cost becomes prohibitively high in large-scale scenarios involving multiple target types, making real-time simulations impractical.

To address these challenges, this study introduces a unified RCS modeling framework that simplifies the representation of scattering behaviors across different target types. By effectively integrating both large-scale and small-scale scattering characteristics, this framework reduces the complexity of modeling and facilitates the standardization of ISAC channels. To validate the proposed framework, comprehensive monostatic RCS measurements were conducted in a large anechoic chamber, focusing on typical ISAC targets. The main contributions of this work are summarized as follows:

\begin{itemize} 
\item[$\bullet$] A unified RCS modeling framework is proposed, which can flexibly adapt to different target types, avoiding fragmented modeling and reducing simulation complexity.  The model consists of three independent components: (i) the large-scale power factor, which quantifies the overall scattering intensity, (ii) the small-scale angular-dependent component, which describes directional scattering behavior, (iii) the random component, accounting for variations between target instances. 

\item[$\bullet$] A parameterization method for the scattering characteristics of different targets, based on measurements at various frequencies and angles, is proposed for the unified RCS framework. 

\item[$\bullet$] Mono-static RCS measurements are conducted for three typical targets (UAV, human, and vehicle) across five central frequencies, with measurements taken over the full azimuth angle range from $-180^{\circ}$ to $180^{\circ}$ at a $5^{\circ}$  step. The experimental results validate the accuracy of the proposed model, which is then integrated into the ISAC channel simulation platform to analyze the impact of different target RCS characteristics on channel properties. \end{itemize}

The remainder of this paper is organized as follows: Section II discusses the modeling approach for the ISAC channel and provides an overview of the current demand for RCS modeling in standardization efforts.  It reviews the theoretical definition of RCS and existing RCS models, summarizes relevant insights, and based on these, proposes a unified RCS modeling framework aimed at meeting the requirements for both large-scale and small-scale channel modeling. Section III describes the measurement setup, experimental procedures, data processing methods, and the unique scattering characteristics of each target derived from the measurement results. Section IV focuses on the parameterization methodology of the proposed RCS model, evaluates its accuracy, and outlines the simulation steps for incorporating RCS into ISAC channel modeling. Additionally, the impact of different targets on channel characteristics is explored through the ISAC channel simulation platform. Finally, Section V summarizes the research findings of this paper.

\section{ISAC channel model and RCS modeling framework}

This section first introduces the modeling approach for the overall ISAC channel, which has gained some consensus in the current 3GPP standardization meetings, and leads into the discussion of the current demand for RCS modeling in the standardization process. On this basis, the theoretical definition of RCS is introduced and a review of existing RCS models is provided, including those for typical regular objects and general RCS calculation methods for more complex and irregular objects that require sophisticated electromagnetic equations. Based on these insights, a unified RCS modeling framework is proposed from the perspective of channel modeling requirements, outlining the necessity of these factors and their physical and mathematical explanations.


\subsection{The RCS modeling requirements for the ISAC channel model}
With several 3GPP ISAC standardization meetings held, there is general agreement to use 3GPP TR 38.901 as the starting point to initiate the standardization work \cite{ref:116meeting}.

According to the 5G 3D GBSM model, assuming that transmitter (Tx) and receiver (Rx) are equipped with arrays of \( S \) and \( U \) antennas, respectively, the channel coefficient from the \( s \)-th antenna of the Tx to the \( u \)-th antenna of the Rx is expressed as
\begin{equation}
	h_{u,s}^{\text{5G}}(t,\tau) = \sum_{n=1}^{N} \sum_{m=1}^{M} h_{u,s,n,m}^{\text{5G}}(t) \cdot \delta(\tau - \tau_{n,m}),
\end{equation}
where \( N \) and \( M \) represent the number of clusters and the number of paths within each cluster, respectively. The channel coefficient \( h_{u,s,n,m}^{\text{5G}}(t,\tau) \) stands for the \( m \)-th ray within the \( n \)-th cluster from the \( s \)-th antenna of the transmitter to the \( u \)-th antenna of the receiver.

\begin{figure*}[!t]
	\centering
	\includegraphics[width=7 in]{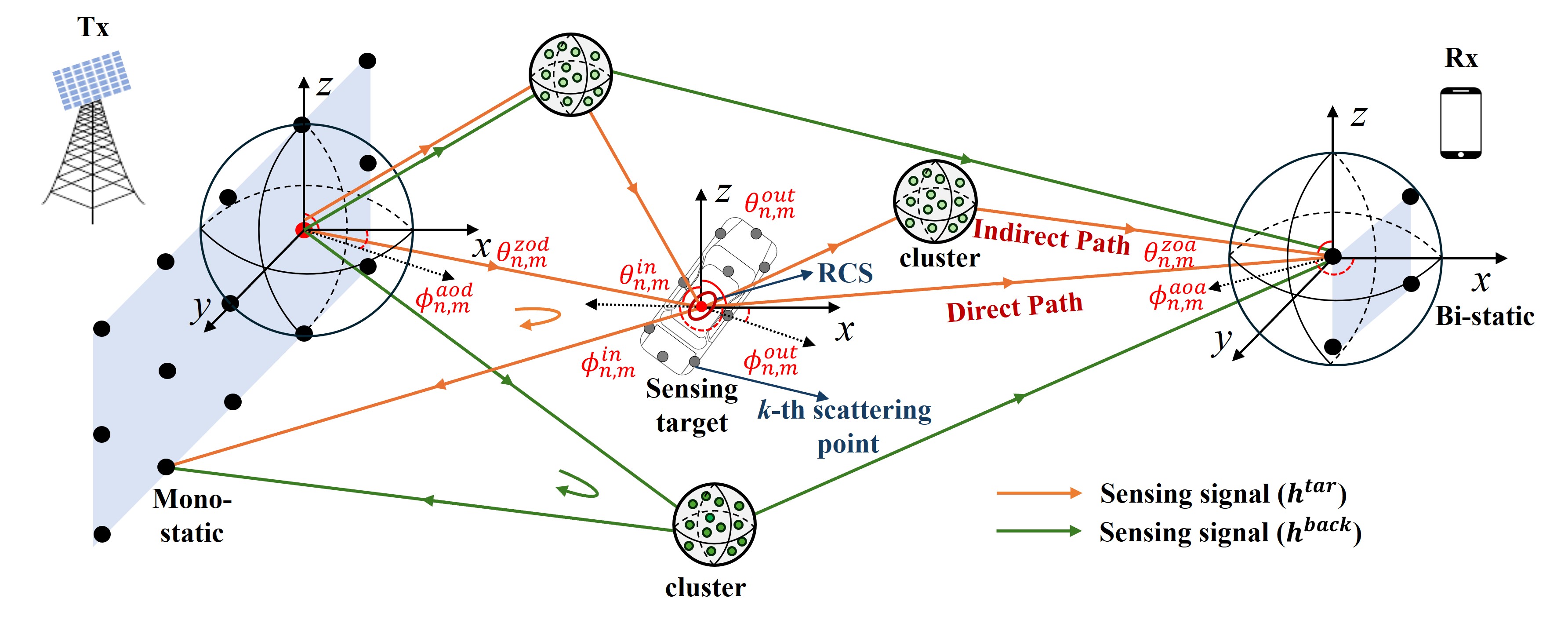}
	\caption{Schematic diagram of 6G ISAC channel model based on GBSM\cite{ref:peinews}.}
	\label{fig:ISAC整体特性大图}
\end{figure*}

Based on this, a consensus was reached at the meetings that the sensing channel is divided into two parts: the target channel and the background channel, as shown in Equation \eqref{eq:感知信道模型}.
The target channel includes all propagation paths influenced by the target, represented by yellow lines in Fig.~\ref{fig:ISAC整体特性大图}. The background channel comprises paths not affected by the target, such as environmental clutter, shown in green lines in Fig.~\ref{fig:ISAC整体特性大图}.

\vspace{-10pt}
\begin{align}
	h^{\text{sen}}_{u,s}(t, \tau) = \sum_{k=1}^K h_{u,s,k}^{\text{tar}}(t, \tau) + h_{u,s}^{\text{back}}(t, \tau),
	\label{eq:感知信道模型}
\end{align}

For the target channel, we assume a total of \(K\) scattering points. These scattering points may originate from a large target that requires multi scattering points modeling, or from multiple distinct targets. For the \(k\)-th scattering points, its channel coefficient can be represented by
\begin{align}
	h_{u,s,k}^{\text{tar}}(t, \tau) &= \sum_{n_1,m_1}^{N_1,M_1} \sum_{n_2,m_2}^{N_2,M_2} h_{u,k,n_2,m_2}^{\text{ST-Rx}}(t) h_{u,k,n_1,m_1}^{\text{Tx-ST}}(t) \notag \\
	& \quad \cdot  \sqrt{\sigma_k(\theta_{\text{out},n_2,m_2},\phi_{\text{out},n_2,m_2},\theta_{\text{in},n_1,m_1},\phi_{\text{in},n_1,m_1})} \notag \\
	& \quad \cdot \exp \left( \textnormal{j} \frac{2 \pi}{\lambda} \left(\mathbf{r}_{\text{in},n_1,m_1}^{\text{T}}\!\cdot\!\mathbf{d}_{\text{sp},k} 
	+ \mathbf{r}_{\text{out},n_2,m_2}^{\text{T}}\!\cdot\!\mathbf{d}_{\text{sp},k} \right) \right) \notag \\
	& \quad \cdot \delta (\tau - \tau_{n_1,m_1}) * \delta (\tau - \tau_{n_2,m_2}),
	\label{eq:目标信道模型}
\end{align}

where,
\begin{itemize}
	\item $n_i,m_i,i\in \left\{ 1,2 \right\}$ indicate the identifier of cluster and path in Tx-ST and ST-Rx links.
	
	\item $h_{u,k,n_1,m_1}^{\text{Tx-ST}}(t)$ and $h_{u,k,n_2,m_2}^{\text{ST-Rx}}(t)$ represent the channel coefficients of the Tx-ST and ST-Rx links, respectively, and their definitions are similar to that of $h_{u,s,n,m}^{\text{5G}}(t)$.
	
	\item  In the ST-Rx link, the Angle of Departure (AoD) of the \((n_2, m_2)\)-th path is characterized by the Azimuth Angle of Departure (AAoD) \(\phi_{\text{out},n_2,m_2}\) and the Elevation Angle of Departure (EAoD) \(\theta_{\text{out},n_2,m_2}\) in a three-dimensional space. Similarly, in the Tx-ST link, the Angle of Arrival (AoA) of the \((n_1, m_1)\)-th path is described by the Azimuth Angle of Arrival (AAoA) \(\phi_{\text{in},n_1,m_1}\) and the Elevation Angle of Arrival (EAoA) \(\theta_{\text{in},n_1,m_1}\).
	
	\item \( \mathbf{r}_{\text{out},n_2,m_2}, \mathbf{r}_{\text{in},n_1,m_1} \) represent the unit direction vectors corresponding to the angles of departure and arrival of the target, respectively. Correspondingly, \( \mathbf{d}_{\text{sp},k} \) represents the position vectors of the $k$-th scattering point of the target.
	
	\item $\sigma_k$ represents the RCS of the $k$-th scattering point of the target, which is dependent on the target itself, as well as the angles of arrival and departure.
\end{itemize}

\begin{table*}[ht]
\centering
\caption{Evaluation Parameters for Different Target Scenarios}
\renewcommand{\arraystretch}{2.5}
\small
\tabcolsep=0.06cm
\begin{tabular}{c|c|c|c|c|c}
\hline
\hline
\textbf{Targets} & \textbf{Automotive} & \textbf{Hazardous Object} & \textbf{Human} & \textbf{AGV} & \textbf{UAV} \\
\hline
Scenarios & \makecell[c]{Highway, \\ Urban grid} & \makecell[c]{Highway, Urban grid, \\ High-speed train} & \makecell[c]{Indoor office, Indoor \\ factory, Indoor room, \\ UMi, UMa, RMa} & InF  & \makecell[c]{UMi, UMa, RMa, \\ UMi-AV, UMa-AV, RMa-AV}  \\
\hline
\makecell[c]{Target \\ Distribution} & \makecell[c]{Based on \\ TR37.885}  & \multicolumn{3}{c|}{Uniformly distributed in a horizontal plane within a defined range} & \makecell[c]{ Uniformly distributed between 1.5 \\ and 300 m in vertical plane \\ or at specific fixed heights}  \\
\hline
Type/Size & \makecell[c]{Passenger car,  \\ Truck/Bus} & \makecell[c]{Child, Animal,  \\ Adult pedestrian} & Child, Adult pedestrian & \makecell[c]{Option 1: 0.5 $\times$ \\ 1.0 $\times$ 0.5 m, Option2: \\ 1.5 $\times$ 3.0 $\times$ 1.5 m}  & \makecell[c]{Option 1: 1.6 $\times$ \\ 1.5 $\times$ 0.7 m, Option2: \\ 0.3 $\times$ 0.4 $\times$ 0.2 m} \\
\hline
\hline
\end{tabular}
\label{tab:ISAC场景}
\end{table*}

It is evident that the ISAC channel model needs to consider the RCS of targets at different angles. Currently, 3GPP standardization efforts focus on five types of sensing targets \cite{ref:ISACScenarios}, as shown in Table~\ref{tab:ISAC场景}: automotives, hazardous objects, humans, automated guided vehicles (AGVs), and UAVs. Each type of target is associated with unique sensing scenarios. The differences in these scenarios determine the geometric layout of the sensing Tx and Rx. In typical scenarios such as Urban Micro (UMi) and Urban Macro (UMa), since targets are usually uniformly distributed in the horizontal plane and the variation in pitch angle is small, the RCS of targets under different azimuth angles is primarily considered. However, in emerging scenarios such as UMi-AV and UMa-AV, where airb orne receivers may be involved, the RCS of targets at different pitch angles must also be taken into account. Additionally, there are variations in the size and distribution of different targets, which directly impact the RCS. Therefore, developing a unified RCS model is crucial for evaluating sensing performance under different target conditions.

\subsection{ Conventional RCS Definitions and Existing Models}
The concept of RCS has been established in the radar domain for decades. The standard definition states that RCS is the hypothetical area required to intercept the incident power at a target such that if the intercepted total power is re-radiated isotropically, it would produce the observed power density at the receiver. Based on this definition, the mathematical expression for monostatic RCS can be readily derived.  Suppose that when propagating in free space, the incident power density $S_I$ at the sensing target can be expressed as

\begin{equation}
     S_{I}=\frac{P_{\text{Tx}} G_{\text{Tx}}}{4 \pi d^{2}} ,
    \label{eqq1}
\end{equation}
where $P_{\text{Tx}}$ is the transmitted power, $G_{\text{Tx}}$ is the Tx antenna gain, $d$ is the distance between the sensing Tx and the sensing target. Therefore, when the sensing Tx and sensing Rx are configured at the same location, the received power $P_{\text{Rx}}$ caused by the isotropic backscatter from the sensing target can be expressed as

\begin{equation}
     P_{\text{Rx}}=\frac{P_{\text{TAR}}}{4 \pi d^{2}} A_{e}=\frac{S_{I} \sigma}{4 \pi d^{2}} A_{e} ,
    \label{eqq2}
\end{equation}
where $P_{TAR}$ provides the amount of power intercepted by the sensing target given $\sigma$ is the effective scattering area (monostatic RCS) and $A_{e}$ is the effective area of Rx antenna. From Equation (\ref{eqq2}), it is straightforward to see that the scattered power density $S_S$ at the Rx is

\begin{equation}
    S_S=\frac{S_I \sigma}{4 \pi d^2}.
    \label{eqq3}
\end{equation}

From Equation (\ref{eqq3}), the monostatic RCS is determined by the following ratio \cite{REF11}

\vspace{-5pt}
\begin{equation}
    \sigma=\lim _{d \rightarrow \infty}\left(4 \pi d^{2} \frac{S_{S}}{S_{I}}\right)=\lim _{d \rightarrow \infty}\left(4 \pi d^{2} \frac{\left|E_{S}\right|^{2}}{\left|E_{I}\right|^{2}}\right),
    \label{eqq4}
\end{equation}
where $E_{S}$ and $E_{I}$ are the far field scattered and incident electric field intensities, respectively.  It is important to emphasize that the RCS defined by Equation (\ref{eqq4}) depends solely on the sensing target itself and is independent of external factors such as path loss, distance, or antenna patterns. Although RCS tends to be larger for larger targets, it does not have a simple linear relationship with the physical size of the target. Instead, RCS can vary significantly depending on several factors, including the target's shape, orientation, material composition, carrier frequency, polarization, and the relative positions of the transmitter and receiver in relation to the geometry of the target.

However, obtaining the RCS of an arbitrary target, as described by Equation (\ref{eqq4}), requires knowledge of the scattered electric field $E_s$. For real-world objects with complex geometries, such as humans, vehicles, and UAVs, determining the scattered field intensity is challenging, making it difficult to compute the RCS directly. To address this, large targets are typically decomposed into smaller facets, each of which is located in the far-field region of both the transmitter and receiver. Under the assumption that there are no mutual interactions between these facets, the total RCS can be determined as the vector sum of the contributions of the individual facets \cite{REF_duomian}

\begin{equation}
\sigma=\left|\sum_{i=1}^{M} \sqrt{\sigma_{i}} e^{j2k \vec{r} \cdot { \vec{c}_i}} e^{j \phi_{i}}\right|^{2},
\label{eqqq555}
\end{equation}
here, $M$ denotes the number of facets, $\sigma_{i}$ represents the RCS of the $i$-th facet,  $\vec{r}$ represents the unit scattering vector directed from the target to the radar, $\vec{c}_i$ is the position vector for the $i$-th  scattering center, and $\phi_i$ is the phase relative to the first facet. The facets can take various simple geometric shapes, such as squares and triangles, depending on the decomposition method used. If the RCS of each facet is known, the total RCS of the target can be obtained by incoherently summing the contributions from all facets.

For well-defined and geometrically simple targets, the RCS can be theoretically determined by solving Maxwell's equations under appropriate boundary conditions, which rigorously describe the interaction between electromagnetic waves and the target. Analytical solutions exist for simple geometries such as spheres, cylinders, ellipsoids, flat rectangular plates, and triangular plates.

The interaction between electromagnetic waves and a target is primarily influenced by the size of the target relative to the wavelength.  This relationship determines the scattering behavior and how RCS is affected by the target's geometry, material properties, and the frequency of the incident wave.  RCS is typically categorized into three regions-optical, Rayleigh, and resonant-based on the dimensionless size parameter $ka=\frac{2 \pi a}{\lambda}$, where $a$ represents the characteristic size of the target and $\lambda$ is the wavelength.

At GHz-level frequencies, most practical targets fall into the optical region $ka\gg 1$, where the target size is much larger than the wavelength.   In this region, scattering can be approximated by using high-frequency methods such as GO, PO, or Geometrical Theory of Diffraction (GTD).

\begin{figure}[!t]
\centering
\includegraphics[width=3.2 in]{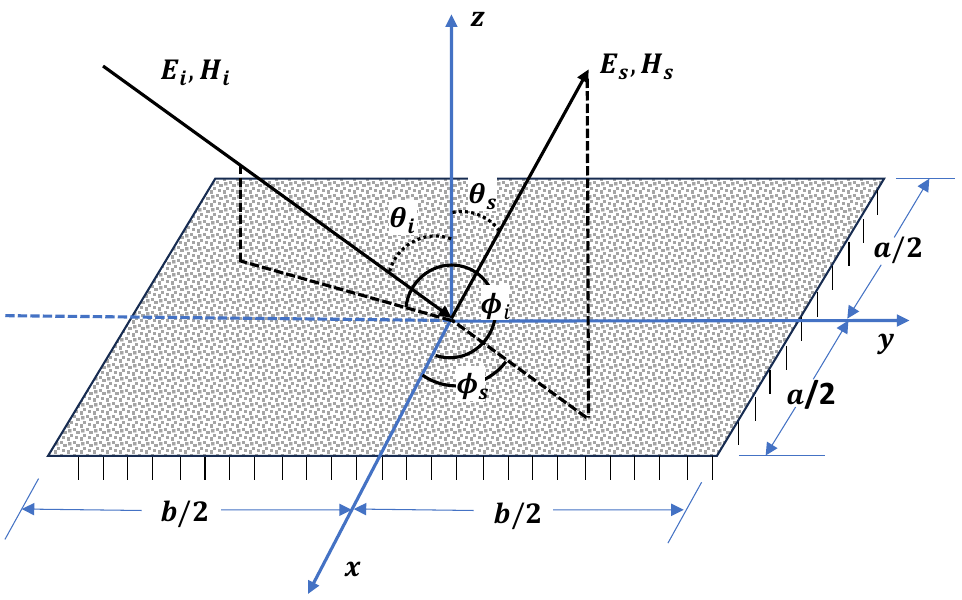}
\caption{Diagram of a uniform plane wave incident on a rectangular conductive plate.}
\label{pingban}
\end{figure}

The rectangular plate serves as the basis for high-frequency scattering analysis. Its simple geometry allows for an analytical RCS solution using the PO approximation while capturing key scattering behaviors.   Consider a perfectly conducting rectangular plate with dimensions $2a$ and $2b$, where $a$ and $b$ are the semilengths of the plate along the $x$- and $y$-axes, as shown in Fig.~\ref{pingban}. A plane wave is incident on the plate at angle $(\theta_i,\phi_i)$, and the scattered field is observed in the direction $(\theta_s,\phi_s)$.

According to the PO approximation, the scattering field of the rectangular plate is given by

\vspace{-5pt}
\begin{equation}
E_{s} \propto E_{i} \frac{e^{-j k r}}{r} 
\iint_{S} e^{j k \left( \sin \theta_{i} \cos \phi_{i} - \sin \theta_{s} \cos \phi_{s} \right) x} \notag
\end{equation}
\begin{equation}
\quad \times e^{j k \left( \sin \theta_{i} \sin \phi_{i} - \sin \theta_{s} \sin \phi_{s} \right) y} \, dS,
\label{eqqq_Es}
\end{equation}
where $k=2\pi / \lambda$ is the wave number, and $r$ is the distance to the observation point. In the backscattering direction $(\theta_i=\theta_s, \ \phi_i=\phi_s=\frac{3 \pi}{2})$, the angular terms simplify. Substituting this into the RCS definition, we obtain the closed-form expression for the RCS

\vspace{-5pt}
\begin{equation}
\sigma_{\textbf{plate}}=4 \pi\left(\frac{a b}{\lambda}\right)^{2} \cos ^{2} \theta_{i}\left[\frac{\sin \left(\beta b \sin \theta_{i}\right)}{\beta b \sin \theta_{i}}\right]^{2}. 
\label{eqqq4}
\end{equation}
 
Table \ref{tab:RCS_Trend_Decomposition} summarizes the RCS expressions for other regular shapes in the optical region, including spheres, cylinders, ellipsoids, cones, and triangular reflectors. These simple shapes are fundamental models for understanding scattering behavior. The RCS of these targets is influenced by several factors, such as the observation angle, operating frequency, target geometry, and material properties. Small changes in any of these factors can lead to significant variations in RCS, complicating accurate modeling. It's also important to note that closed-form RCS solutions for these targets are typically derived under specific conditions, such as in the optical region or for particular incident and scattering angles. Under more general conditions, the RCS of simple targets may become a complex function, making it challenging to accurately characterize their scattering behavior with simplified models.

\begin{table*}[ht]
\centering
\caption{RCS Expressions and Trend Decomposition for Common Canonical Shapes}
\renewcommand{\arraystretch}{2.5}
\small
\tabcolsep=0.45cm
\begin{tabular}{c|c|c|c}
\hline
\hline
\textbf{Shape} & \makecell[c]{\textbf{RCS Expression}\\ \textbf{(Optical Region)}} & \makecell[c]{\textbf{Overall Scattering}\\ \textbf{Capability (Factor 1)}} & \makecell[c]{\textbf{Angular Dependency}\\ \textbf{(Factor 2)}} \\ 
 \hline

Sphere & \( \displaystyle \sigma = \pi a^2 \) & \( \displaystyle \pi a^2 \) & \( \displaystyle 1 \) (Isotropic) \\ \hline
Ellipsoid & \( \displaystyle \sigma = \frac{\pi a^2 b^2}{\lambda^2} \frac{1}{\left( \cos^2\theta + \frac{a^2}{b^2} \sin^2\theta \right)^2} \) & \( \displaystyle \frac{\pi^2 b^3(a^2+b^2)}{a\lambda^2} \) & \( \displaystyle \frac{a^3}{\pi{b^3}(1+\frac{a^2}{b^2} ) } \frac{1}{\left( \cos^2\theta + \frac{a^2}{b^2} \sin^2\theta \right)^2} \) \\ \hline
Cone & \( \displaystyle \sigma = \frac{\pi a^2}{\lambda^2} \frac{\sin^2(\theta - \alpha)}{(\theta - \alpha)^2} \) & \( \displaystyle \frac{\pi a^2}{\lambda^2}*P \) & \( \displaystyle \frac{1}{P}\frac{\sin^2(\theta - \alpha)}{(\theta - \alpha)^2} \) \\ \hline
Triangle Reflector & \( \displaystyle \sigma = \frac{4\pi}{\lambda^2} \frac{a^4}{(h + b)^2} \cos^2\theta \) & \( \displaystyle \frac{4\pi^2}{\lambda^2} \frac{a^4}{(h + b)^2} \) & \( \displaystyle \frac{1}{\pi}\cos^2\theta \) \\ \hline
\hline
\end{tabular}
\label{tab:RCS_Trend_Decomposition}

\vspace{0.5em} 
\begin{flushleft}
\textbf{Note:} 
\(a, b\): Semi-length and semi-width of the flat plate. 
\(L\): Length of the cylinder. 
\(k\): Wave number. 
\(\theta\): Angle of incidence or observation. 
\(\alpha\): Cone angle. 
\(h, b\): Dimensions of the triangle reflector.
And $P=(\text{Si}(4\pi-2\alpha)+\text{Si}(2\alpha)+\text{sin}^2(\alpha)*(1/(2\pi-\alpha) - 1/\alpha))$, $\text{Si}(x)$ denotes the sine integral.
\end{flushleft}
\end{table*}

Existing models often struggle to systematically characterize RCS due to its inherent complexity. However, a closer analysis of these expressions reveals two distinct factors in RCS behavior. First, we integrate the RCS over $\theta$  and normalize by $2\pi$,  defining this value as the overall scattering capability. Second, the RCS normalized by the overall scattering capability is termed the angular dependency. The overall scattering capability quantifies the total scattered power, which depends on the target's size and the wavelength, while the angular dependency describes how the RCS varies with changes in the observation or incident angle, influenced by phase differences and interference effects. Consequently, we can simplify the RCS description as a function of wavelength $k$ and incident angle $\theta$, i.e., $\sigma(f,\theta)$.  This insight is foundational for the unified modeling framework proposed in this paper.

\subsection{Proposed Unified RCS Modeling Framework}

Currently, existing RCS models typically focus on specific target types, making it difficult to provide a unified representation for a wide range of targets. Moreover, considering that different target types exhibit different scattering patterns, modeling each target type independently significantly increases the complexity of channel modeling and ISAC system simulation. For complex targets, detailed electromagnetic simulations are often required, and the computational complexity grows with the complexity of the target’s geometry and material properties. These challenges motivate the need for a generic and flexible model that can accommodate different target types while balancing computational efficiency and modeling accuracy.

We propose a unified RCS modeling framework, represented as

\vspace{-5pt}
\begin{equation}
\sigma (f, \phi) = A(f) \times B(f,\phi),
    \label{eq_IIII_1}
\end{equation}
where $f$ denotes the center frequency, and $\phi$ represents the azimuth angle. The RCS is decomposed into two independent components to meet the requirements of the channel model.     The first component, $A(f)$,   represents the large-scale scattering power, quantifying the target's average ability to intercept and scatter electromagnetic waves. It is normalized in the angular domain and is expressed as

\begin{equation}
A(f) = \frac{1}{2\pi} \int_{0}^{2\pi} \sigma(f,\phi) d\phi.
    \label{eq_IIII_1}
\end{equation}

Thus,  $A(f)$ is independent of angular variations and serves as a key input to the large-scale path loss model. It is determined by the target's physical properties, such as size and material composition, and primarily captures the frequency dependence of the RCS. The scattering capability at different frequencies often exhibits significant variations, as described by $A(f)$.

The second component, $B(f,\phi)$, represents the angular dependence, describing how the RCS varies with incident and observation angles.  This component captures the target's directional scattering characteristics, strongly influenced by its shape and orientation.  Different targets exhibit unique angular patterns, crucial for modeling small-scale directional effects in ISAC systems.  For instance, flat surfaces produce strong specular reflections, while cylindrical or irregular targets exhibit more complex angular behaviors.  The inclusion of $B(f,\phi)$ enables the framework to accurately model scattering directionality for various target types.

In practical scenarios, even variations within the same target type, such as large and small vehicles or people in different clothing, can lead to significant fluctuations in RCS.  This variability arises from numerous factors, such as surface roughness, environmental conditions, and multipath effects, making RCS a sensitive parameter.  To address this, we introduce a third component, $B_2$,  to account for these fluctuations

\begin{equation}
\sigma (f, \phi) = A(f) \times B_1(f,\phi)  \times B_2.
    \label{eq_IIII_1}
\end{equation}

Component $B_2$ captures random variations and is typically modeled using statistical distributions like log-normal, Weibull, or Gamma.   This is particularly important for dynamic and irregular targets in real-world environments.   The inclusion of  $B_2$ ensures that the model accounts for the inherent randomness in scattering behavior, enhancing its accuracy and flexibility.

\section{RCS Measurements and Observations}
\subsection{Measurement Setup}
\begin{figure*}[ht]
\centering
\subfloat[]{\includegraphics[width=1.73in]{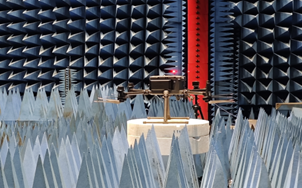}
\label{mental1}}
\hfill
\subfloat[]{\includegraphics[width=1.73in]{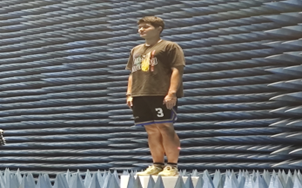}
\label{mental2}}
\hfill
\subfloat[]{\includegraphics[width=1.73in]{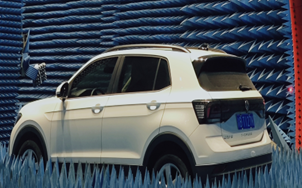}
\label{mental3}}
\hfill
\subfloat[]{\includegraphics[width=1.73in]{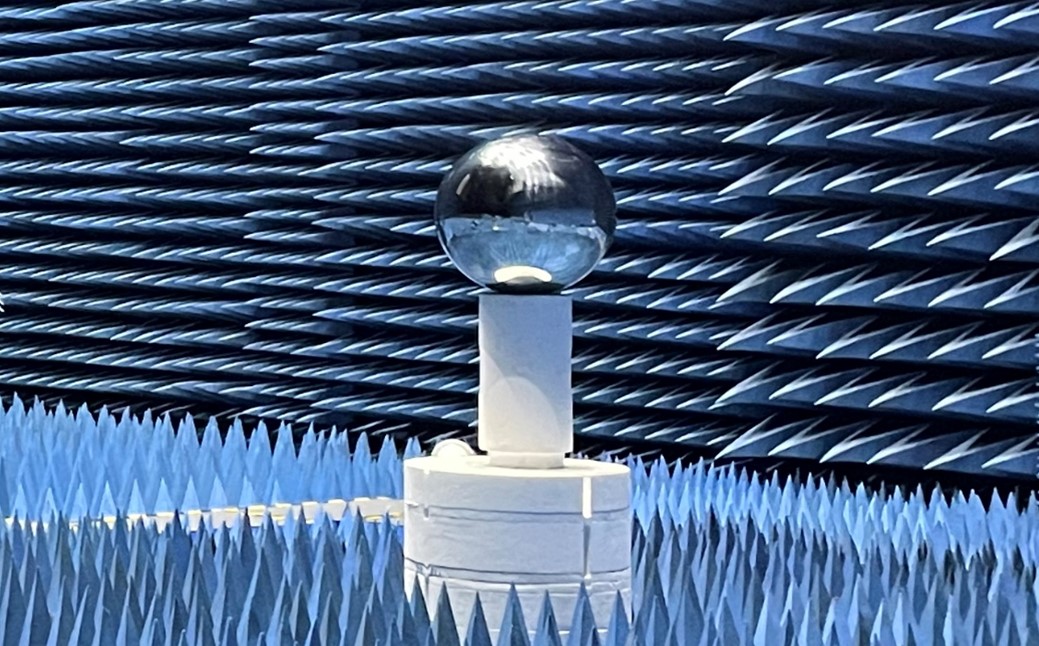}
\label{mental4}}

\captionsetup{ font=small, width=\textwidth}
\caption{RCS measurement scenarios of three typical targets in an anechoic chamber. (a) UAV. (b) Human. (c) Vehicle. (d) Metal sphere.}
\label{imag_1}
\end{figure*}

The anechoic chamber effectively minimizes environmental interference, providing an ideal setting for measuring the RCS of various targets, including UAVs, humans, and vehicles, as shown in Fig. \ref{imag_1}. The UAV used is the DJI M350, with four symmetrically deployed rotor arms and approximate dimensions of 430 × 420 × 430 mm. The human model represents an individual with a height of 180 cm and weight of 70 kg. The vehicle is a Volkswagen T-Cross model, measuring 4.218 × 1.76 × 1.599 m, and the metallic sphere has a diameter of 0.5 meters.


\begin{figure}[!t]
\centering
\includegraphics[width=3.2 in]{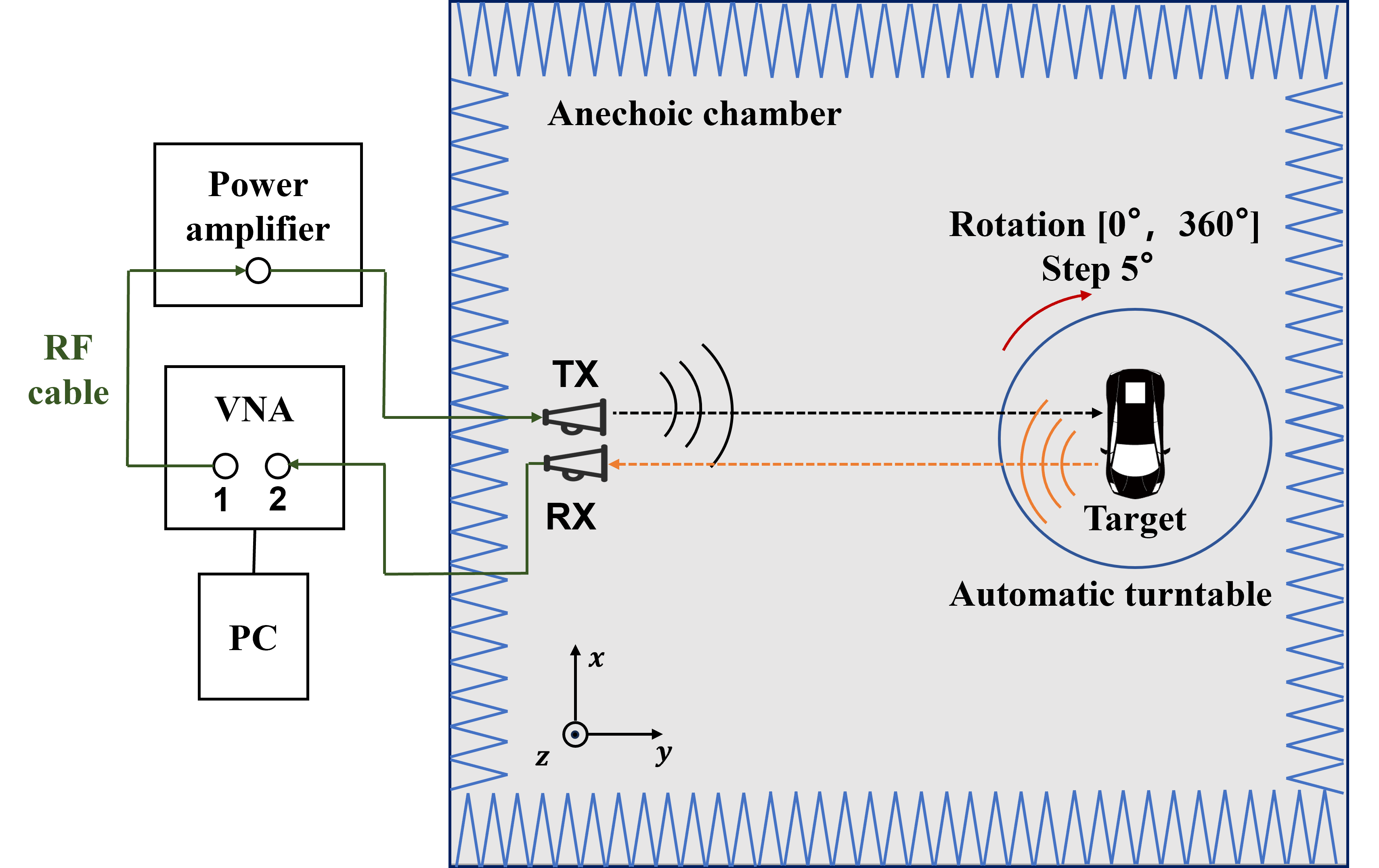}                
\caption{The diagram of VNA measurement platform.}
\label{imag_2}
\end{figure}

The measurement system consists primarily of a vector network analyzer (VNA), power amplifiers, two horn antennas (to transmit and receive sensing signals), a PC controller and other auxiliary components, as shown in Fig. \ref{imag_2}. The VNA serves as the core device, transmitting frequency-stepped signals and receiving and storing frequency domain response signals. The two horn antennas are positioned approximately 10 cm apart to simulate monostatic sensing. The sensing target is placed in the center of an automated turntable, which rotates $360^{\circ}$ in increments of $5^{\circ}$. This allows measurement of the target's echoes at various angles. It should be noted that this study focuses on two-dimensional RCS measurements in the horizontal plane, with the elevation angle fixed at $90^{\circ}$.


As detailed in Table \ref{table_1}, the measurement procedure consists of two main parts. First, RCS measurements are conducted for UAVs, humans, and vehicles across five carrier frequencies (10 GHz, 15 GHz, 20 GHz, 28 GHz, and 36 GHz), aiming to assess the scattering characteristics of different targets at various angles under monostatic sensing. Second, a 0.5-meter-diameter metal sphere is used as a calibration target to ensure the accuracy of the RCS measurements and subsequent data analysis. The sphere is placed on a foam support to minimize measurement errors.

\begin{table}[htbp] 
\centering
\caption{Measurement Configuration}
\label{table_1}
\small 
\setlength{\tabcolsep}{3pt} 
\renewcommand{\arraystretch}{1.2} 
\begin{tabular}{c|c|c|c|c}
\hline
\hline
\textbf{Measurement Content} & \multicolumn{3}{c|}{\textbf{\makecell{RCS Modeling \\ with Angle}}} & \textbf{\makecell{Validation of \\ Data Processing}} \\ \hline
Target Type & UAV & Human & Vehicle & Metal Sphere  \\ \hline
Frequency (GHz) & \multicolumn{3}{c|}{36  /28 /20 /15 /10 } & 28  \\ \hline
\makecell{Number of \\ Frequency Samp    les}  & \multicolumn{4}{c}{801} \\ \hline
Bandwidth (GHz) & \multicolumn{4}{c}{3} \\ \hline
\makecell{Distance between \\ Tx/Rx and Target (m)} & 6 & 12 & 12 & 6  \\ \hline
Target Rotation Angle & \multicolumn{3}{c|}{[0, $360^{\circ}$], Step $5^{\circ}$} & [0, $360^{\circ}$], Step $5^{\circ}$  \\ \hline
Tx Antenna Gain (dBi)  & \multicolumn{3}{c|}{13 / 25} & 25  \\ \hline
Rx Antenna Gain (dBi) & \multicolumn{3}{c|}{13 / 25} & 25  \\ \hline
3dB Beamwidth  & \multicolumn{3}{c|}{$34^{\circ}$ / $10^{\circ}$} & $10^{\circ}$  \\ \hline
Polarization & \multicolumn{4}{c}{Vertical} \\ \hline
Sensing Mode & \multicolumn{3}{c|}{Monostatic} & Monostatic  \\ \hline
\hline
\end{tabular}
\end{table}

\subsection{Measurement procedure and data processing}
During the measurement process, the VNA collects the transmission coefficient $|S_{21}|$, which characterizes the transmission properties between the Tx and Rx. This coefficient reflects the combined effects of the two-link propagation paths (Tx-ST and ST-Rx) and the target's scattering response. To isolate the target's scattering characteristics, the radar equation is employed

\begin{equation}
     \sigma = \frac{P_r (4\pi)^3 R_1^2 R_2^2}{P_t G_t G_r \lambda^2},
    \label{eq1_rcs}
\end{equation}
where $\sigma$ denotes the RCS , $P_r$ is the received power, $P_t$ is the transmitted power, $G_t$ and  $G_r$ are the gains of Tx and Rx antennas,  $R_1$ and $R_2$ are the distances from the target to Tx and Rx, and $\lambda$ is the wavelength.

\begin{figure}[!t]
\centering
\includegraphics[width=3.2 in]{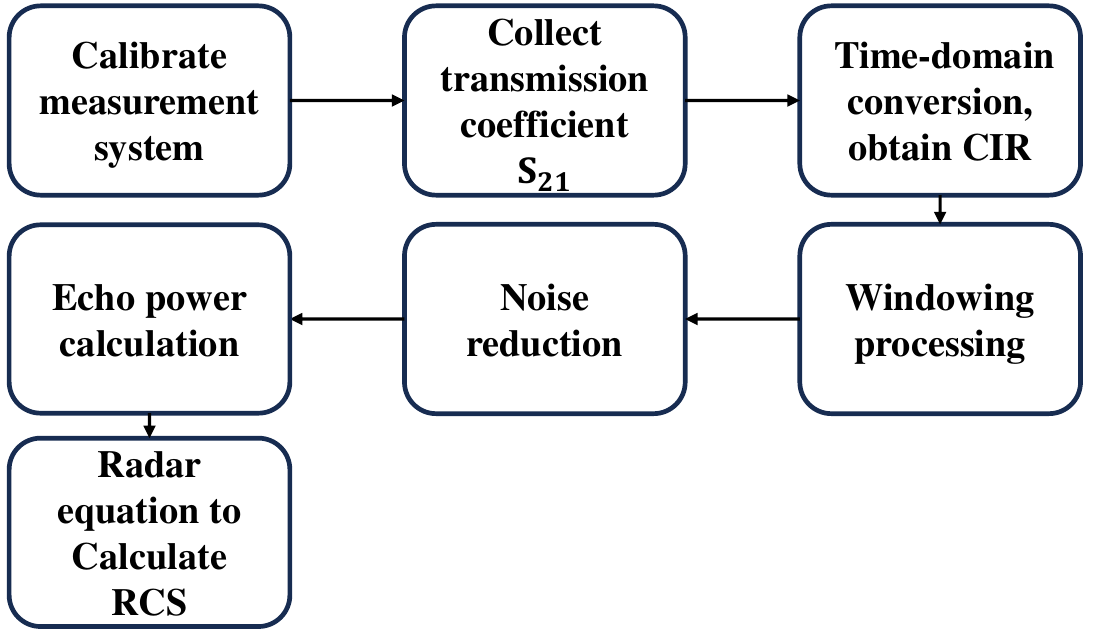}
\caption{Illustration of data post-processing.}
\label{imag_3}
\end{figure}

To ensure accurate results, the data undergoes additional processing. Fig. \ref{imag_3} illustrates the workflow. After system calibration to eliminate equipment-induced imperfections, the  $|S_{21}|$ data, collected in the frequency domain, are converted to the time domain. The Power Delay Profile (PDP) identifies multipath components based on signal delays, while windowing and denoising techniques are applied to mitigate interference. The Time-Domain CIR is then used to calculate echo power by summing the squared signal amplitudes. Finally, the results are inserted into Equation (\ref{eq1_rcs}) to compute the RCS.

\begin{figure}[!t]
\centering
\includegraphics[width=3.2 in]{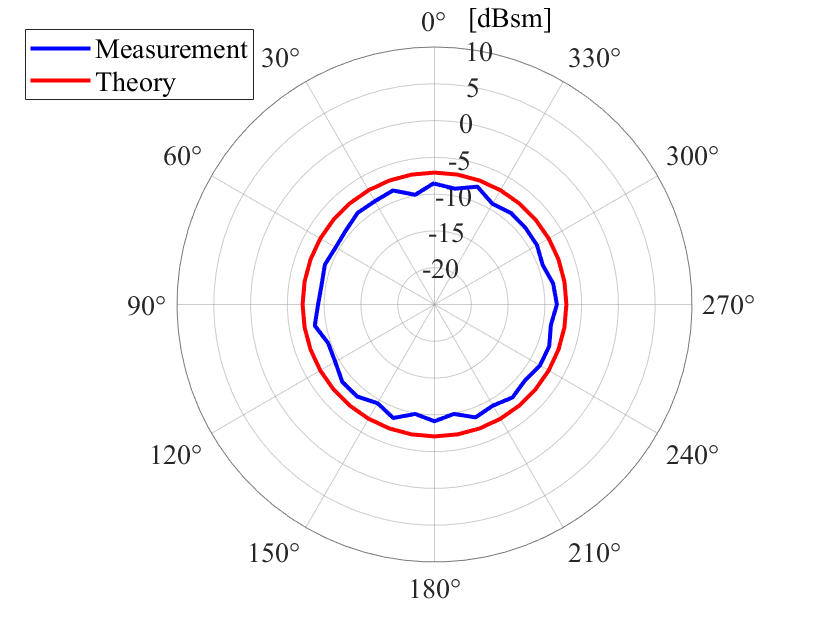}
\caption{Measured vs. Theoretical RCS of the Metal Sphere with a diameter of 0.5 m. }
\label{imag_5}
\end{figure}

A 0.5-meter-diameter metal sphere is used as a calibration target. With its size much larger than the 28 GHz wavelength, it operates in the geometric optical region. The theoretical RCS is derived from Table \ref{tab:RCS_Trend_Decomposition}, and a comparison of the measured and theoretical RCS values is shown in Fig. \ref{imag_5}. The theoretical RCS is -7.07 dBsm, while the measured average is -8.96 dBsm, with a discrepancy of less than 2 dBsm, confirming the high accuracy of the measurements. The slight difference is attributed to surface roughness and non-ideal reflective properties of the sphere. These results validate the experimental setup and provide a foundation for further analysis.

\subsection{Experimental Observations: measurement results for different targets}

\begin{figure*}[ht]
\centering
\subfloat[]{\includegraphics[width=2.31in]{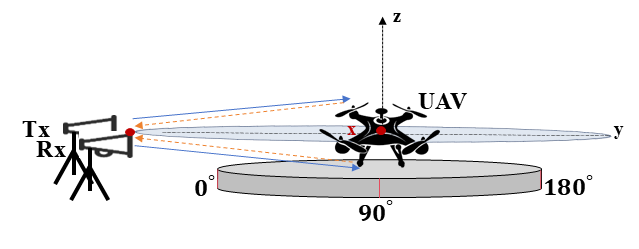}
\label{uav10}}
\hfill
\subfloat[]{\includegraphics[width=2.31in]{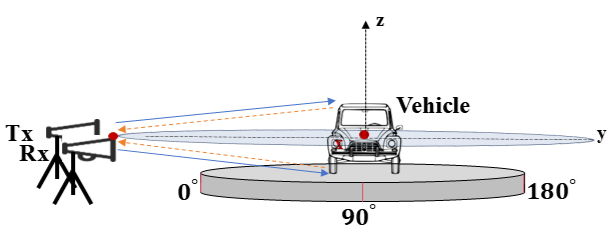}
\label{car10}}
\hfill
\subfloat[]{\includegraphics[width=2.31in]{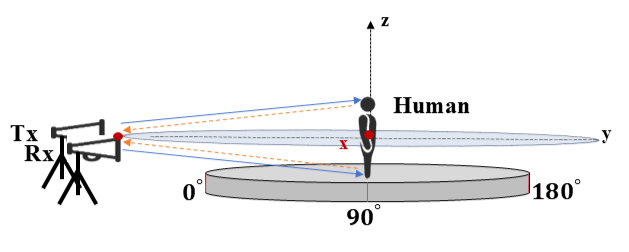}
\label{human10}}
\hfill

\subfloat[]{\includegraphics[width=2.31in]{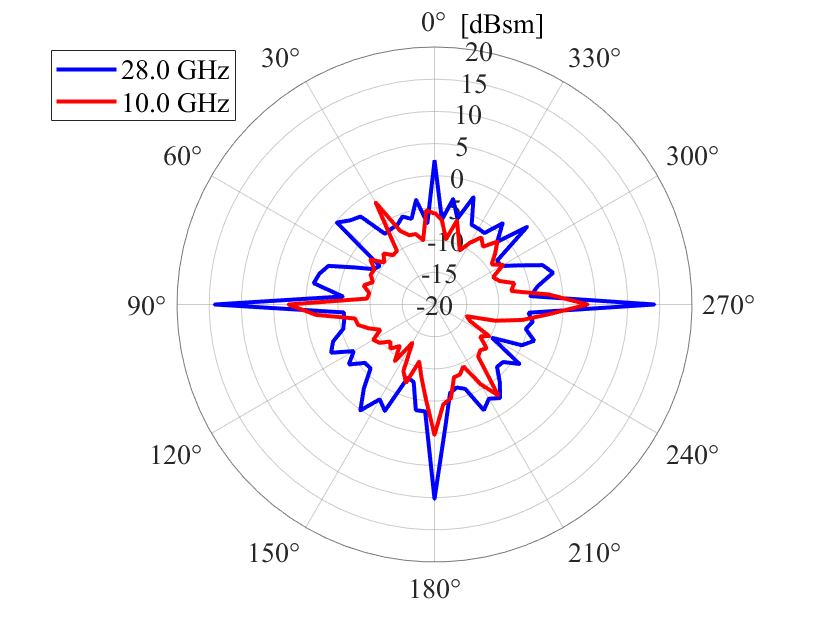}
\label{uav28}}
\hfill
\subfloat[]{\includegraphics[width=2.31in]{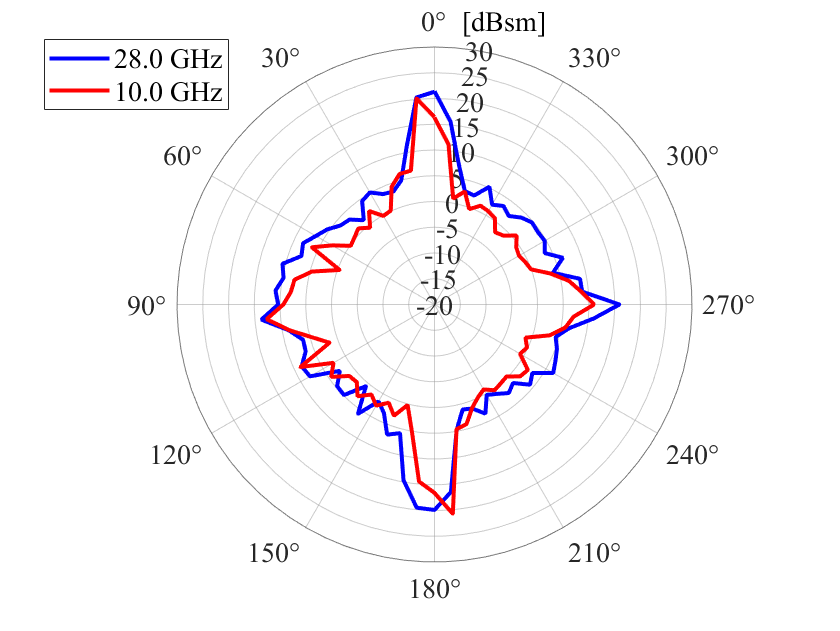}
\label{car28}}
\hfill
\subfloat[]{\includegraphics[width=2.31in]{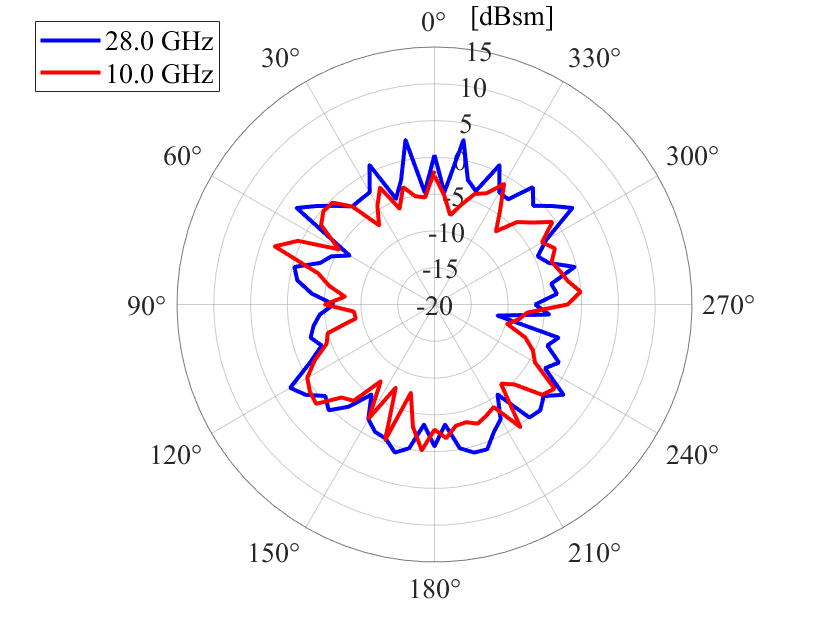}
\label{human28}}

\captionsetup{font=small, width=\textwidth}
\caption{
RCS of three targets versus azimuth angle at 10 GHz and 28 GHz.
(a), (b), and (c) are the measurement schematics for UAV, Vehicle, and Human, respectively;
(d), (e), and (f) are the corresponding measurement results at 10 GHz and 28 GHz, respectively.
}
\label{imag_all22}
\end{figure*}

Based on the RCS measurements conducted in the anechoic chamber for three typical targets in the previous section, this section presents the RCS results at 10 GHz and 28 GHz, used as representative frequencies due to space constraints. Fig. \ref{imag_all22} shows schematic diagrams (a), (b), and (c) to help readers identify the corresponding scattering characteristics at different angles. The measurement results for the three targets are shown in Figs.~\ref{imag_all22}(d), (e), and (f). The following section analyzes the observed scattering characteristics for each target.

a) UAV

As shown in Fig. \ref{imag_all22}(d), the UAV exhibits higher RCS values around $0^{\circ}$, 90$^{\circ}$, $180^{\circ}$, and $270^{\circ}$, forming a four-leaf shape. This is due to the UAV's rectangular body shape, with most electromagnetic waves being reflected back to the receiving antenna from these directions, resulting in higher RCS at these angles. At other angles, the RCS values are lower, showing almost omnidirectional scattering with some fluctuations..

b) Vehicle

The RCS of the vehicle, shown in Fig.  \ref{imag_all22}(e),  follows a similar pattern to the UAV, with larger RCS values around 0$^{\circ}$, 90$^{\circ}$, 180$^{\circ}$, and 270$^{\circ}$. However, there are notable differences compared to the UAV. The peak widths and RCS values differ; for the vehicle, each peak has a width of approximately 30$^{\circ}$, which is slightly broader than that of the UAV. Furthermore, the average RCS of the vehicle is higher than that of the UAV. This is attributed to the vehicle's significantly larger size compared to the UAV under identical testing conditions, providing a greater reflective surface area for electromagnetic waves, resulting in stronger echoes received at the antenna.

c) Human

As shown in Fig. \ref{imag_all22}(f), the human body exhibits a scattering pattern that can be approximated as isotropic, with minimal angular dependence but some random fluctuations.

Based on these results, we observe that the scattering characteristics of the UAV and vehicle show angular dependence, whereas the human body has minimal angular dependence.  This difference is due to the geometric properties of the targets: the human body is approximately cylindrical, maintaining a consistent reflective area, while the UAV and vehicle are similar to rectangular prisms, reflecting signals more strongly when facing their sides.

A comparison of the average RCS values across the frequency bands reveals the following trend: vehicle\textgreater human\textgreater UAV.  This aligns with expectations, as the average RCS is largely influenced by the physical size of the target.  Additionally, as the frequency increases, the average RCS of all targets rises significantly.  This notable finding will be discussed further in the next section.

\section{Model parameterization and validation}
This section presents the parameterization method for the unified RCS model, validates the model, and integrates it into the ISAC simulation platform to analyze RCS channel characteristics for different targets.

\subsection{Parameterization Methodology}
In the proposed unified RCS model, the RCS of a sensing target is decomposed into three main components.  For any type of target, the first component, $A(f)$, represents the average intensity of the electromagnetic energy intercepted by the target and scattered back to the transmitter.  This component is angle-independent and is typically obtained by calculating the linear average of the RCS values measured at all angles, as defined below

\vspace{-5pt}
\begin{equation}
A(f)=\frac{\sum_{i=1}^{N} \sigma_{i}(f,\phi_i)}{N},
    \label{eq_IIII_model_1}
\end{equation}
where $N$ represents the number of angles, and  $\sigma_{i}(f,\phi_i)$ denotes the RCS value at the $i$-th angle.  Component $A(f)$ is frequency-dependent, exhibiting significant variations with different carrier frequencies. Based on measured results, we assume that $B_1$ remains relatively stable with frequency, so we parameterize it as $B_1(f, \phi) = B_1(\phi)$.

The second component, $B_1(\phi)$, can be understood as analogous to an antenna radiation pattern, representing the relative gain of scattering at different incident and observation angles. $B_1(\phi)$ is strongly influenced by the shape of the target’s reflective surface. From the RCS patterns of the three targets analyzed in the previous section, $B_1(\phi)$ is generated using two approaches.

For targets with shapes that approximate spheres or cylinders (e.g., the human body), the scattering pattern is predominantly isotropic, so an angle-dependent model is not necessary. In these cases, $B_1(\phi)$ is set to 1 across all directions. However, for targets with significant angular dependence (e.g., UAVs and vehicles, which have rectangular shapes), $B_1(\phi)$ must be modeled as a function of angle, expressed as $B_1(\phi) = F(\theta=90^\circ, \phi)$, where $\theta$ and $\phi$ are the elevation and azimuth angles, respectively.

To highlight the specific significance of $B_1(\phi)$, the RCS value at each angle is normalized by dividing it by $A(f)$, i.e., $\frac{\sigma_i}{A(f)}$, to represent the relative gain at different angles. This normalization removes the influence of large-scale power levels and isolates the angular characteristics of the target's scattering behavior. The normalized data is then used to fit and derive $B_1(\phi)$.

For UAV and vehicles, we observe significant RCS peaks around $0^\circ$, $90^\circ$, $180^\circ$, and $270^\circ$, with each peak exhibiting a shape that resembles a quadratic function. In other angular regions, the RCS approximates an omnidirectional pattern. For targets with strong angular dependence, we propose modeling $B_1(\phi)$ using a piecewise function, as defined in 3GPP antenna radiation power pattern specifications \cite{3GPPPPPP}. This method allows precise characterization of the four distinct beam peaks. The function uses quadratic approximations in peak angular regions and constant values in off-peak domains, as formalized below


\vspace{-5pt}
\begin{equation}
10 \log_{10} B_{1}\left(\theta = 90^\circ, \phi \right) = -\min \left\{
\begin{split}
    &12\left(\frac{\phi - \phi_{k}}{\phi_{\text{3dB},k}}\right)^{2} + c_{k}, \\
    &Y_{\max }
\end{split}
\right\},\\
\label{eq_UAV_B1}
\end{equation}
where $\phi_k,\phi_{\text{3dB},k},c_k$ denotes the angular center, the 3dB beamwidth and the peak gain of the $k$-th beam peak, respectively. Within the mainlobe region, the quadratic term dominates with smaller values, forcing the minimum operator to select the beam peak expression. As angular deviation $|\phi-\phi_k|$ increases, the quadratic term become larger than $Y_{\max}$, causing the minimum operator to select $-Y_{\max}$ as the value of $B_1$.

In addition to components $A$ and $B_1$, we introduce a stochastic fluctuation factor, $B_2$, which effectively models the randomness and variability of RCS in real world environments. After isolating the large-scale power component $A$ and the small-scale angular component $B_1$ from the RCS data, the remaining term, $\frac{\sigma_i}{A \times B_1}$, typically exhibits irregular behavior. These residual data are used to fit the stochastic fluctuation factor $B_2$, which is modeled using statistical distributions.

To accurately represent the stochastic component $B_2$, this study incorporates three commonly used distributions for modeling the superposition of signal fields: log-normal, gamma, and Weibull distributions. These distributions are widely applied in existing studies and are frequently used to simulate the scattering characteristics of various targets.

\subsection{Model validation}
This section validates the proposed unified model and provides parameter values for different targets.

\begin{figure}[!t]
\centering
\includegraphics[width=3.2 in]{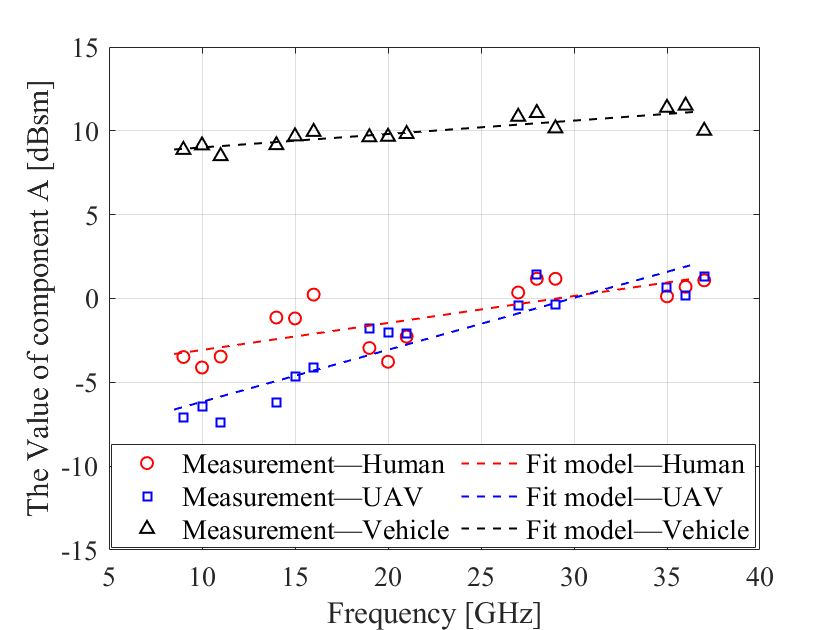}
\caption{Measured and Fitted $A$ for Different Targets (Human, UAV, and Vehicle). }
\label{imag_51}
\end{figure}

We begin by extracting the component $A(f)$ from the RCS data of three targets. RCS measurements were taken at five central frequencies with a 3 GHz bandwidth, divided into 1 GHz increments, resulting in data at 15 frequency points. As $A$ is frequency-dependent, we fit the data for $A$, as shown in Fig. \ref{imag_51}.

\begin{figure*}[ht]
\centering
\subfloat[]{\includegraphics[width=2.3in]{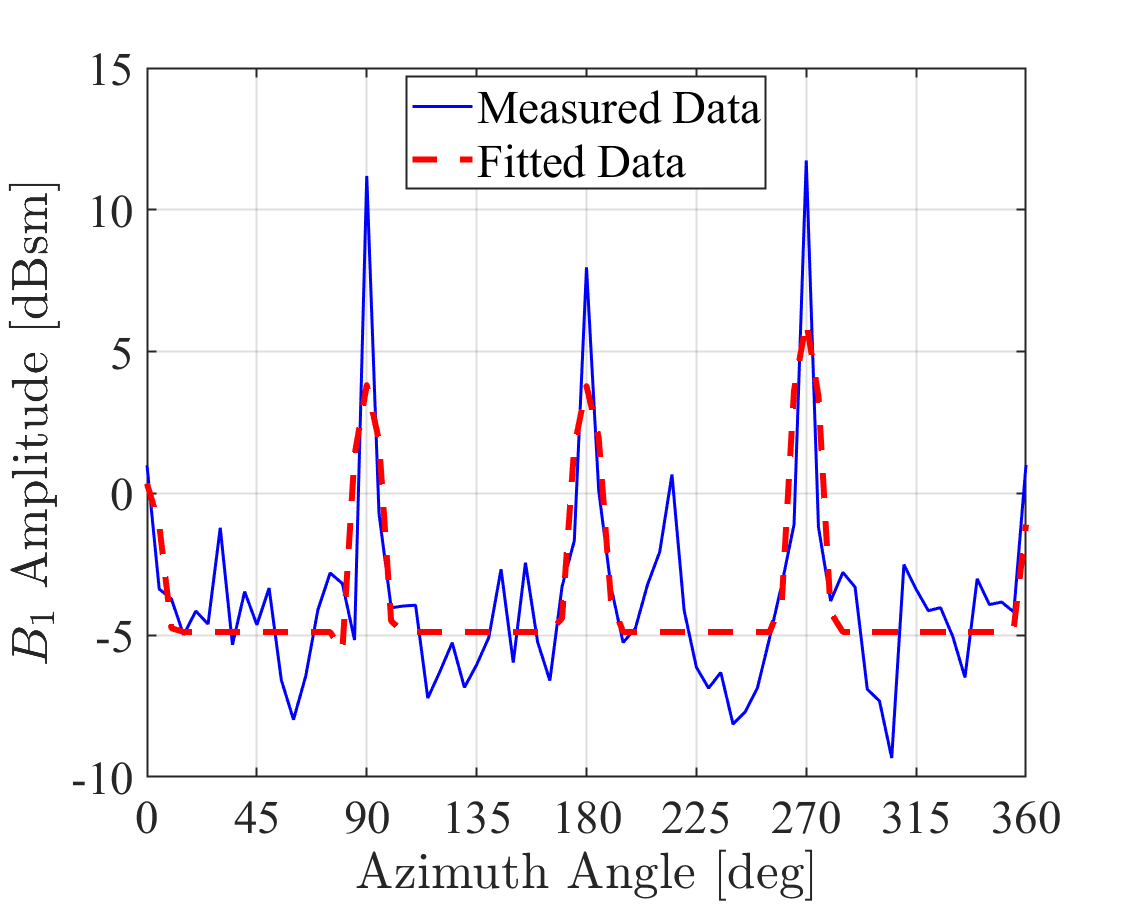}
\label{uav1}}
\hfill
\subfloat[]{\includegraphics[width=2.3in]{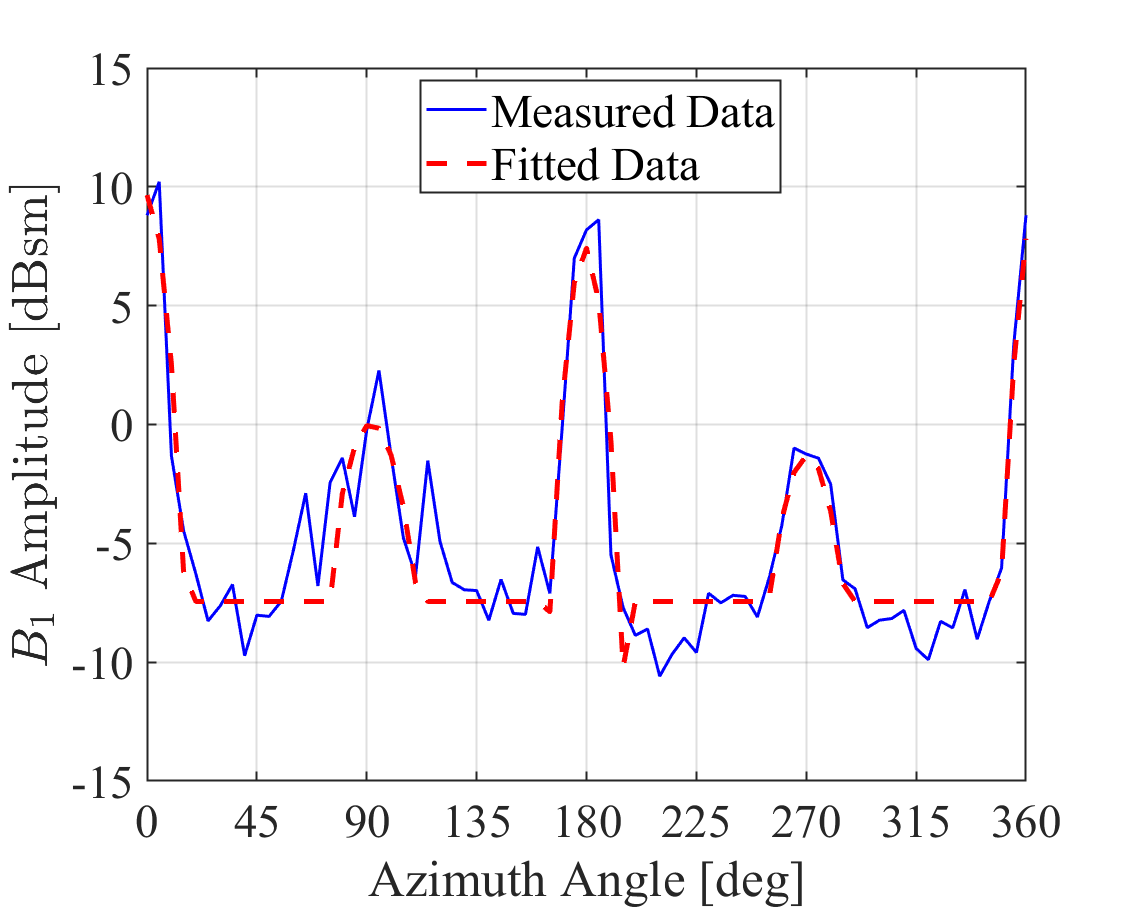}
\label{car1}}
\hfill
\subfloat[]{\includegraphics[width=2.3 in]{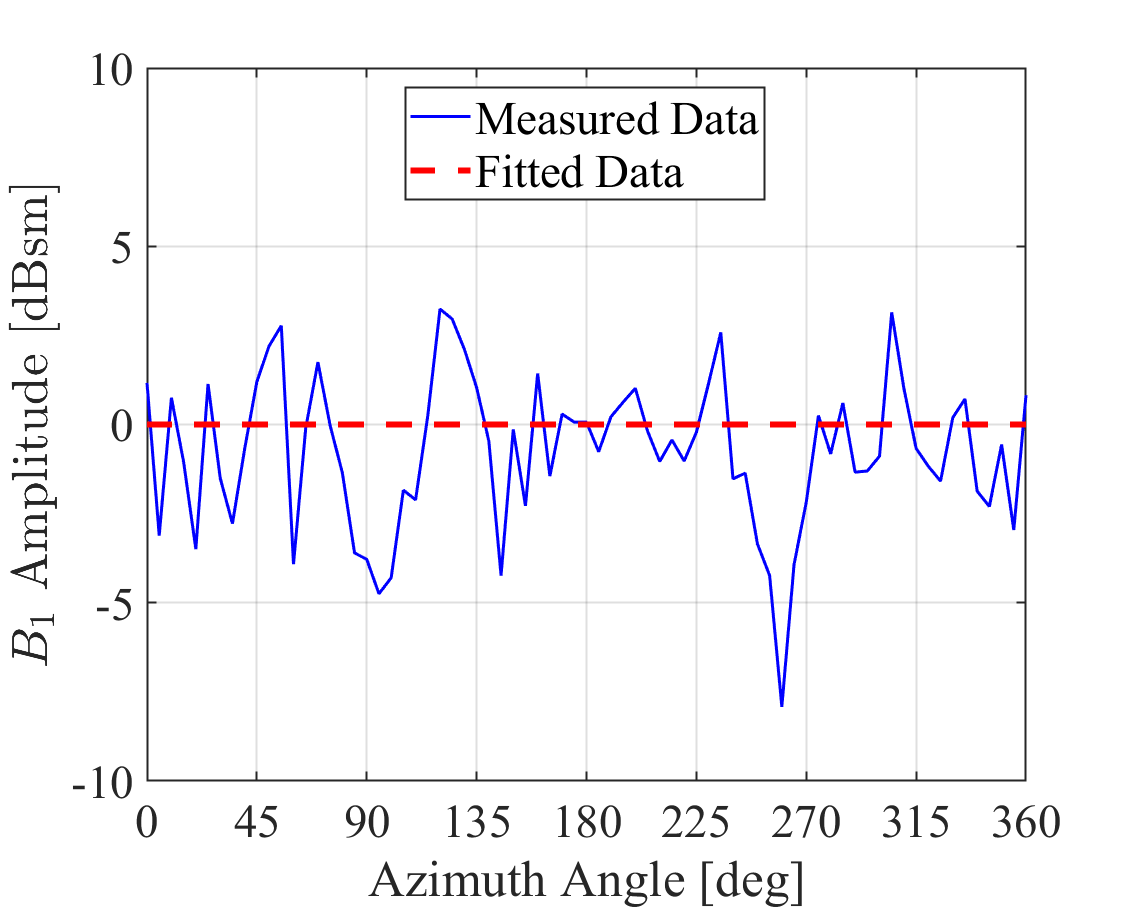}
\label{human1}}

\captionsetup{ font=small, width=\textwidth}
\caption{Comparison of Measured and Modeled Results for $B_1(\phi)$ : (a) UAV, (b) Vehicle, (c) Human. }
\label{rcs_human_car_uav}
\end{figure*}

The RCS values of the three targets fluctuate with frequency, showing a general upward trend. Based on this observation, we model the frequency-dependent component $A$ as a monotonic linear function of frequency. The fitting equations for the three targets are as follows

\begin{equation}
\left\{
\begin{aligned}
A_{\text{UAV}} \ \  &= 0.31 \cdot f - 9.26, \\
A_{\text{vehicle}} &= 0.08 \cdot f + 8.21, \\
A_{\text{human}} &= 0.16 \cdot f - 4.68,      
\end{aligned}
\right.
\label{eq1_f}
\end{equation}
where $f$ represents the frequency.  After extracting $A$, the remaining data are fitted using the proposed $B_1(\phi) $ method. The fitting results are shown in Fig. \ref{rcs_human_car_uav}.

Figs. \ref{rcs_human_car_uav}(a) and (b) demonstrate that the model accurately captures the angular dependence of both the UAV and the vehicle, particularly the distribution of the four prominent peaks. The fitted curves show good accuracy in identifying peak positions and widths, although some deviations are observed in peak amplitudes. For the human target, Fig. \ref{rcs_human_car_uav}(c) shows the $B_1(\phi)$ curve fluctuating irregularly around 0 dB, as expected, without distinct peaks.

The Root Mean Square Error (RMSE) is used to evaluate the model's fitting accuracy. A smaller RMSE indicates higher predictive accuracy. The formula is as follows

\begin{equation}
   \text{RMSE} = \sqrt{\frac{1}{N} \sum_{i=1}^N \left( y_i - \hat{y}_i \right)^2},
    \label{eq1}
\end{equation}
where $y_i$ are the measured values, and $\hat{y}_i$ are the model's predicted values. The RMSE values for the UAV, vehicle, and human are 2.414, 1.8309, and 2.2451, respectively. These results show better fitting performance for the vehicle, especially at peak positions. For the human target, the lack of angular dependency leads to irregular oscillations, and the random fluctuation factor $B_2$ alone effectively models its behavior.

\begin{figure*}[ht]
\centering
\subfloat[]{\includegraphics[width=2.33in]{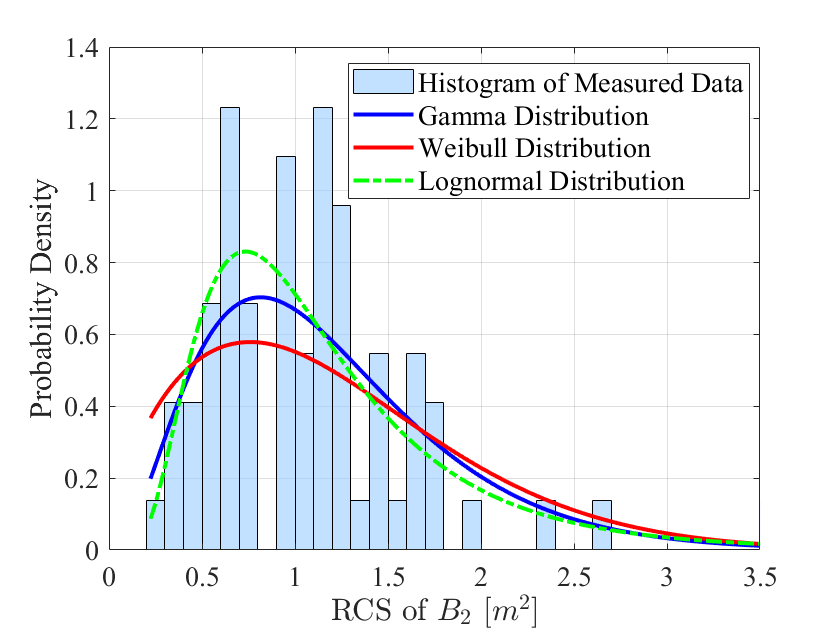}
\label{uav2}}
\hfill
\subfloat[]{\includegraphics[width=2.33in]{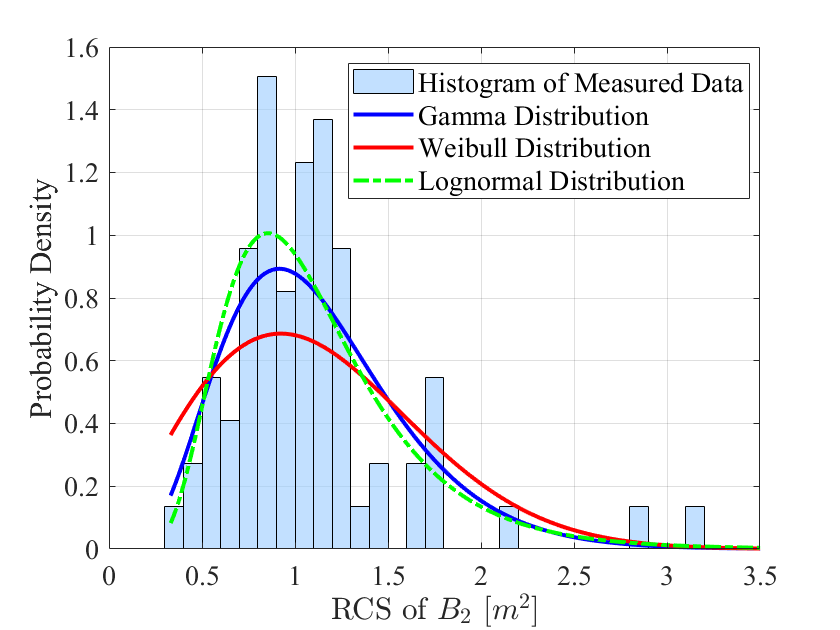}
\label{car2}}
\hfill
\subfloat[]{\includegraphics[width=2.33in]{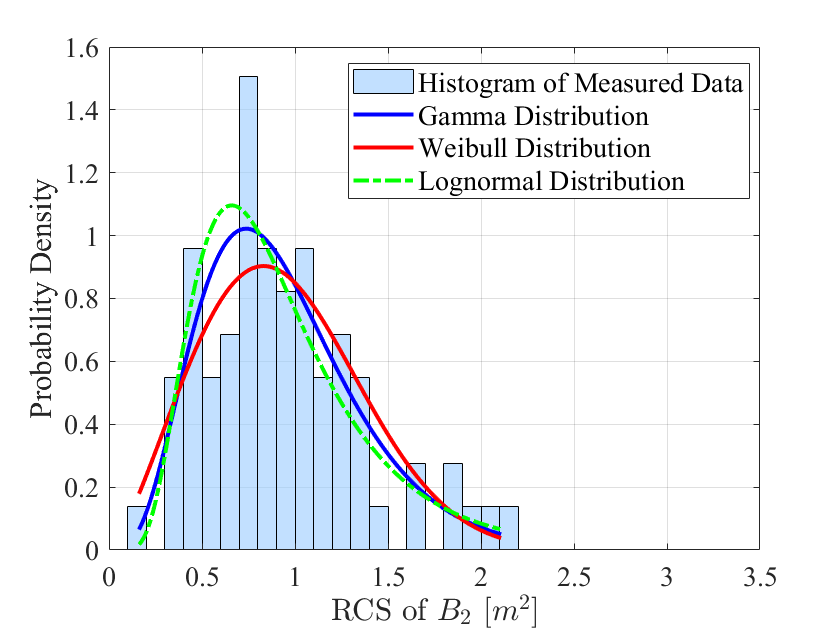}
\label{human2}}
\captionsetup{ font=small, width=\textwidth}
\caption{PDFs of three statistical distributions for three targets: (a) UAV, (b) Vehicle, (c) Human. }
\label{imag_pdf_v1}
\end{figure*}

After extracting the large-scale component $A(f)$ and small-scale component $B_1(\phi)$ from the RCS data, the remaining data typically exhibits irregular patterns, attributed to random fluctuations in the target's RCS. This component is modeled using statistical distribution functions. Fig. \ref{imag_pdf_v1} presents the PDFs for three distributions along with empirical data, with parameters estimated using Maximum Likelihood Estimation (MLE).

To evaluate the performance of fitting of three distribution functions, a Kullback-Leibler (KL) divergence metric \cite{REF211_KL}, defined as

\vspace{-5pt}
\begin{equation}
D_{KL}(p \| q)=\sum_{i=1}^{N} p\left(x_{i}\right) \cdot\left(\log p\left(x_{i}\right)-\log q\left(x_{i}\right)\right),
    \label{eq_UAV_B1}
\end{equation}
where $p_i$ represents the probability of the actual RCS measurement at the $i$-th azimuth angle, and $q_i$ is the fitted distribution probability from the model. A smaller KL divergence indicates a better model fit. Table \ref{table_pdf} presents the fitting results for component $B_2$ using measured data and different distribution functions.

\begin{table}[htbp]
\centering
\caption{KL Divergence for Component $B_2$ with Different Distribution Functions}
\label{table_pdf}
\small 
\setlength{\tabcolsep}{16pt} 
\renewcommand{\arraystretch}{1.5} 
\begin{tabular}{c|c|c|c}
\hline
\hline
\multirow{2}{*}{\textbf{Target}} & \multicolumn{3}{c}{\textbf{KL Divergence}} \\ \cline{2-4}
                & \textbf{Lognormal} & \textbf{Weibull} & \textbf{Gamma} \\ 
\hline
UAV            & 0.0535             & 0.1014           & 0.0693 \\  \hline
Vehicle        & 0.0674             & 0.1676           & 0.1021 \\     \hline    
Human          & 0.032             & 0.0134           & 0.0134 \\ 
\hline
\hline
\end{tabular}
\end{table}

For all three targets, while the divergence values of the log-normal distribution are similar to those of the other two distributions, we ultimately selected the log-normal distribution for modeling the empirical data. This decision is based on two factors: First, the log-normal distribution's parameters ($\mu$ and $\sigma$) have clear physical meanings, corresponding directly to the mean and standard deviation of the normal distribution in logarithmic space. Second, from an engineering perspective, the log-normal distribution offers computational efficiency and analytical properties, which enhance its practical applicability in engineering tasks. Therefore, we opted to model component $B_2$ with the log-normal distribution.


At this stage, we have thoroughly analyzed the modeling and fitting processes for the three components, $A$, $B_1$, and $B_2$, within the overall RCS model framework. Table \ref{all_A1_B1_B2} presents the RCS parameter fitting results for the three target objects measured.

\begin{table*}[ht]
\centering
\newcommand{\tabincell}[2]{\begin{tabular}{@{}#1@{}}#2\end{tabular}}
\caption{The  fitting parameter  results for UAV, vehicle and human}
\renewcommand{\arraystretch}{1.2}
\small
\tabcolsep=0.6cm
\begin{tabular}{
    p{0.7cm}
    |p{3cm}
    |p{1.3cm}
    |p{0.4cm}
    |p{0.7cm}
    |p{1cm}
    |p{2cm}}
\hline
\hline
\multirow{2}{*}{\textbf{Target}} 
&\multirow{2}{*}{\textbf{$\quad\quad\bm{A}$ [dBsm]}} 
&\multicolumn{4}{c|}{\textbf{$\bm{ B_1}$ [dBsm]}}
&\multirow{2}{*}{\textbf{$\quad\bm{B_2}$[dBsm]}}
\\
\cline{3-6}
    & 
    & \textbf{$\ \ \quad\phi\in$} 
    & \textbf{$\phi_k$} 
    & \textbf{$\phi_{3dB,k}$} 
    & \textbf{$\quad\ c_k$} 
    & 
    \\ 
\hline
    \multirow{5}{*}{UAV} 
    &\multirow{5}{*}{$\begin{aligned}&A=-9.26+0.31*f\end{aligned}$}
    & $\begin{aligned}&[315^\circ,360^\circ)\\&\cup[0^\circ,45^\circ]\end{aligned}$ 
    & $\ \ 0^\circ$ 
    & $20.84^\circ$ 
    & $\ \ \ \ \ 0.68$
    & \multirow{5}{*}{$N(-0.52,2.31^2)$} 
    \\
\cline{3-6}
    & 
    & $[45^\circ,135^\circ]$ 
    & $\ 90^\circ$ 
    & $10.47^\circ$ 
    & $\ \ -6.52$
    &
    \\
\cline{3-6}
    & 
    & $[135^\circ,225^\circ]$ 
    & $180^\circ$ 
    & $15.41^\circ$ 
    & $\ \ -5.61$
    &
    \\ 
\cline{3-6}
    & 
    & $[225^\circ,315^\circ]$ 
    & $270^\circ$ 
    & $14.51^\circ$ 
    & $-12.30$
    &
    \\ 
\cline{3-6}
    & 
    & \multicolumn{4}{c|}{$Y_{\max }=4.47$} 
    & 
    \\ 
\hline
    \multirow{5}{*}{Vehicle} 
    & \multirow{5}{*}{$\begin{aligned}&A=8.21+0.08*f\end{aligned}$}  
    & $\begin{aligned}&[315^\circ,360^\circ)\\&\cup[0^\circ,45^\circ]\end{aligned}$ 
    & $\ \ 0^\circ$ 
    & $12.59^\circ$ 
    & $\ \ -9.61$
    & \multirow{5}{*}{$N(-0.53,2.64^2)$} 
    \\
\cline{3-6}
    & 
    & $[45^\circ,135^\circ]$ 
    & $\ 90^\circ$ 
    & $30.42^\circ$ 
    & $\ \ -1.67$
    &
    \\
\cline{3-6}
    & 
    & $[135^\circ,225^\circ]$ 
    & $180^\circ$ 
    & $15.79^\circ$ 
    & $-10.06$
    &
    \\ 
\cline{3-6}
    & 
    & $[225^\circ,315^\circ]$ 
    & $270^\circ$ 
    & $28.03^\circ$ 
    & $\quad\ 1.56$
    &
    \\ 
\cline{3-6}
    &
    & \multicolumn{4}{c|}{$Y_{\max }=7.46$} 
    & 
    \\ 

\hline
Human
& $\begin{aligned}&A=-4.68+0.16*f\end{aligned}$
& \multicolumn{4}{c|}{0} 
& $N(-0.77,2.13^2)$ 
\\ 
\hline
\hline

\end{tabular}%
\label{all_A1_B1_B2}
\end{table*}


\subsection{Implementation Framework and Simulation Results of the Unified RCS Model in ISAC channel}

The proposed unified RCS modeling framework seamlessly integrates into the ISAC channel modeling workflow by building on the foundational 3GPP TR 38.901 channel model \cite{REF21}. This integration introduces a novel enhancement, enabling accurate simulation of the Tx-ST-Rx sensing link alongside the traditional Tx-Rx communication link. 


Fig. \ref{liuchengtu_ISAC} illustrates the updated ISAC target channel modeling process, which primarily involves the integration of RCS-related steps. It is worth noting that only the modeling steps for the sensing target channel are presented here, while the communication channel modeling steps are omitted, as the communication channel does not account for the target's RCS. In the figure, we emphasize the newly added steps related to RCS configuration, detailed as follows:

\begin{figure}[!t]
\centering
\includegraphics[width=3.2 in]{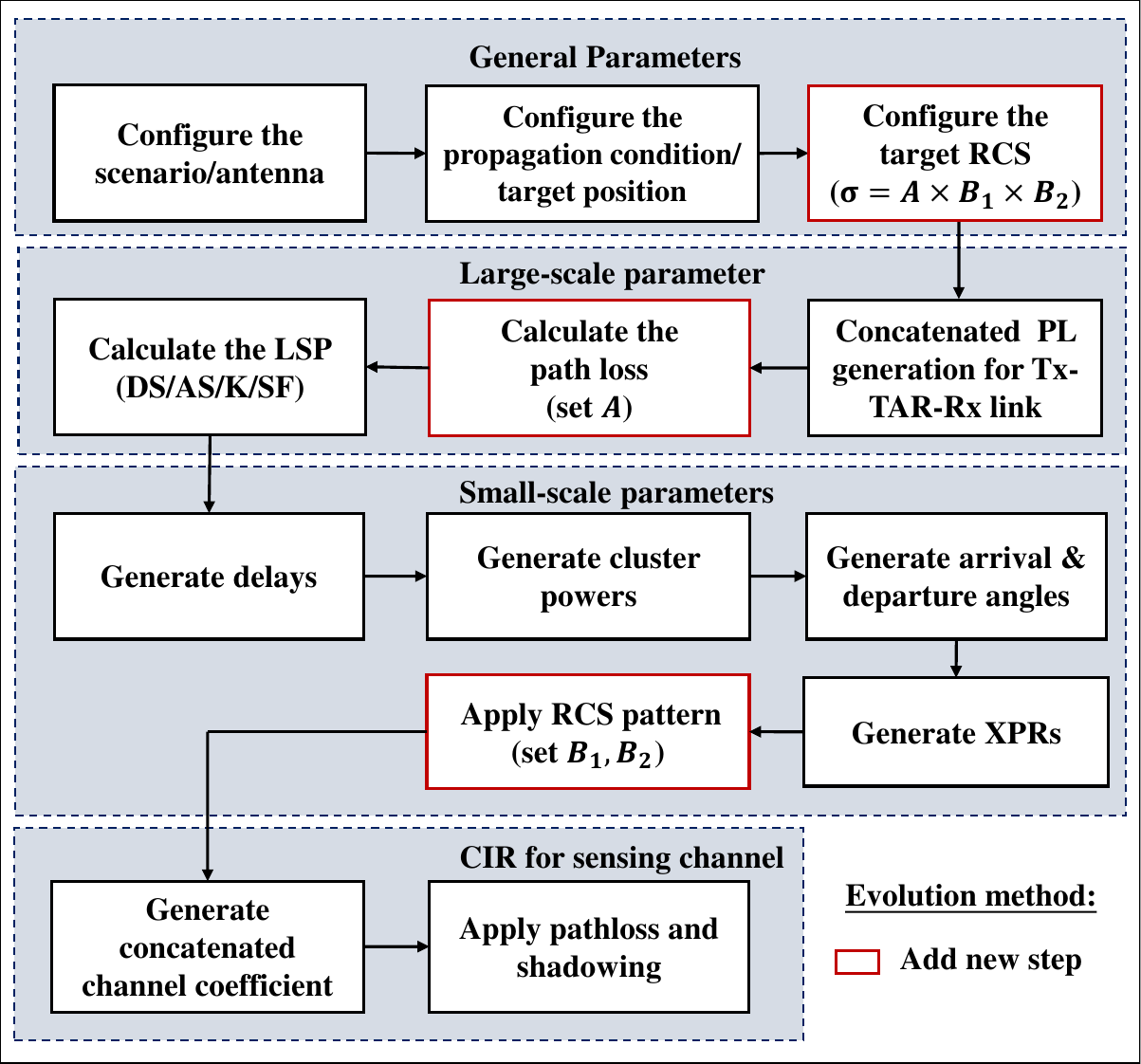}
\caption{The evolution of ISAC target channel simulation under the unified RCS  modeling framework.}
\label{liuchengtu_ISAC}
\end{figure}

\begin{itemize} 
\item[$\bullet$] \textbf{Target RCS Configuration}: Based on the target type and frequency band specified in the simulation configuration, the corresponding RCS statistical parameters are obtained by consulting a lookup table (e.g., Table \ref{all_A1_B1_B2}). The specific values for $A$, $B_1$, and $B_2$ are then determined by sampling from the corresponding distributions.
\item[$\bullet$] \textbf{Target Channel Path Loss Generation}: During the generation of large-scale parameters, the method for generating the target channel path loss is modified. In this context, the RCS model introduces the large-scale scattering power factor ($A$), which determines the overall power level of the scattered signal. A higher value of $A$ corresponds to stronger reflected signals, lower path loss, and higher received power at the receiver. Since large-scale parameters do not account for angular variations or specific multipath effects, only the $A$ component is required at this stage, while the remaining components, $B_1$ and $B_2$, are set to 1.
\item[$\bullet$] \textbf{Apply RCS Pattern}: During the generation process of small-scale parameters, the application of the RCS pattern is incorporated. At this stage, the RCS model uses $B_1$ to capture the angular properties of the target (e.g., geometry and orientation) and their impact on the strength of the multipath components, based on the AOD and AOA. The random fluctuation factor $B_2$ accounts for the variations caused by destructive interference, which introduces stochastic changes to the channel. Therefore, the values of $B_1$ and $B_2$ are the main focus, while the $A$ component can simply be set to 1.
\end{itemize}

The unified RCS model bridges large-scale and small-scale channel modeling by decomposing the RCS into three independent and interpretable components: $A$, $B_1$, and $B_2$. This modular structure enhances generality, computational efficiency, and scalability, enabling the representation of diverse targets, from canonical shapes to complex dynamic objects, within a unified mathematical framework. By separately modeling large-scale scattering, angular dependency, and random variability, the model provides a flexible and efficient tool for ISAC system simulations across various scenarios.

\begin{figure*}[ht]
\centering
\subfloat[]{\includegraphics[width=2.33in]{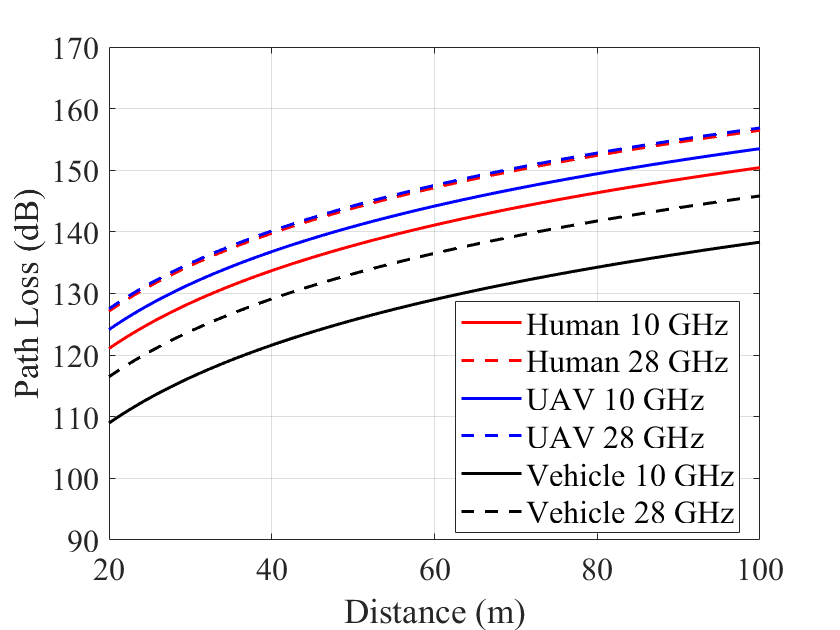}
\label{uav111}}
\hfill
\subfloat[]{\includegraphics[width=2.33in]{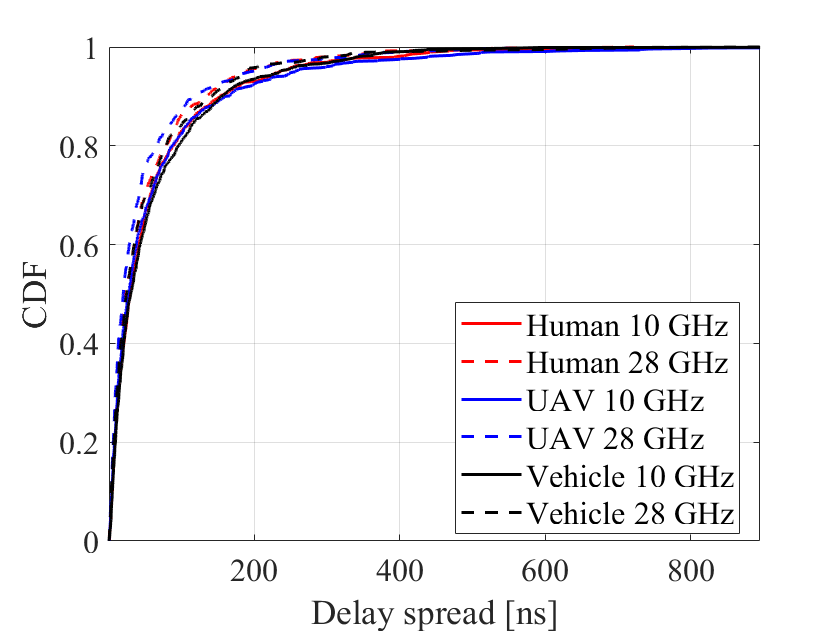}
\label{car111}}
\hfill
\subfloat[]{\includegraphics[width=2.33in]{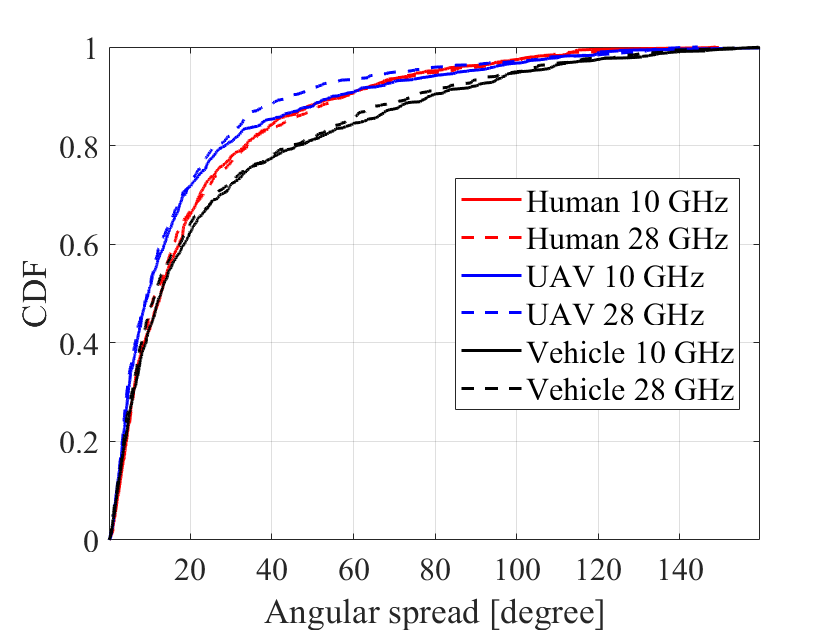}
\label{human111}}
\captionsetup{ font=small, width=\textwidth}
\caption{Simulation results of different channel parameters for three targets at 10 GHz and 28 GHz: (a) path loss, (b) delay spread, (c) angular spread. }
\label{channel_simulation}
\end{figure*}

To validate its effectiveness, the unified RCS model is integrated into the BUPTCMCCCMG-IMT2023 channel simulation platform \cite{REF_pingtai}, which is specifically designed for ISAC channel evaluation. The simulation follows the UMi environment defined in ITU-R M.2412 \cite{REF_ITU}, with a baseline setup consisting of a single base station and user terminal, both equipped with a single antenna. To focus on the impact of target RCS, each simulation considers a single target. This section presents simulation results for large-scale path loss, small-scale delay spread, and angular spread, as shown in Fig. \ref{channel_simulation}, analyzing the influence of different target RCS characteristics on the sensing channel.

1) Path loss

As shown in Fig. \ref{channel_simulation}(a), the RCS of different targets significantly impacts the path loss characteristics of the sensing channel. At the same frequency, the path loss is highest for the UAV, followed by the human, while the vehicle exhibits the lowest path loss. This is because a larger RCS leads to stronger signal reflection, resulting in lower path loss and higher received power. Therefore, targets with higher RCS values tend to achieve better sensing performance under certain conditions.

Additionally, frequency has a notable effect on path loss. High-frequency signals (28 GHz) exhibit much higher path loss compared to low-frequency signals (10 GHz), especially for targets with lower RCS, such as humans, where scattering and absorption effects are more pronounced. As a result, the path loss for humans experiences a greater degree of attenuation when moving from 10 GHz to 28 GHz compared to the UAV.

2) Delay spread and angular spread

As shown in Fig. \ref{channel_simulation} (b) and (c), the RCS characteristics of different targets influence the distribution of delay spread  and angular spread. This is the result of the combined effect of the RCS pattern and the power spectrum.
First, comparing the three targets, the vehicle target exhibits a pronounced directional dependence in its $B_1$ parameter, especially along four distinct propagation directions where the power is significantly enhanced. This leads to a more dispersed multipath distribution, which in turn increases both delay spread and angular spread. In contrast, the RCS of the human and UAV targets shows weaker angular dependence, resulting in a less pronounced effect on the channel's delay spread  and angular spread compared to the vehicle target. Furthermore, the frequency has minimal impact on delay spread and angular spread, as no significant frequency dependence is observed in the $B_1$ measurements at 10 GHz and 28 GHz. The results of delay spread  and angular spread are primarily influenced by the profile shape of the target's RCS, while the $A$ parameter has a relatively minor effect.


\section{Conclusion and future work}
This paper proposes a unified RCS modeling framework for ISAC channel modeling, designed to accurately represent the scattering characteristics of targets and meet the diverse requirements of channel modeling.  Specifically, the framework decomposes the RCS into three independent components, each representing the effects of scattering power, angular dependence, and random fluctuations.  Through experimental measurements and data analysis, the model has been validated to effectively capture the scattering characteristics of different target types, demonstrating its versatility and applicability.  Furthermore, the RCS model has been integrated into the ISAC channel simulation platform, optimizing the ISAC channel modeling process.  Simulation results for large-scale path loss, small-scale delay spread, and angular spread are presented, along with an analysis of how different target RCS characteristics influence the sensing channel.  These findings provide valuable insights and strong theoretical support for the design and optimization of ISAC systems.

In future work, we aim to expand our research on RCS modeling by developing a more comprehensive 3D RCS model, which will improve the accuracy of the modeling, particularly in the context of complex, dynamic targets. This expansion will allow for a more precise representation of RCS across different target orientations and geometries. Additionally, we will focus on exploring dual-station RCS models, which will further enhance the modeling capabilities for multi-sensor ISAC systems.

\bibliographystyle{IEEEtran}
\bibliography{bib}

\begin{thebibliography}{10}
\providecommand{\url}[1]{#1}
\csname url@samestyle\endcsname
\providecommand{\newblock}{\relax}
\providecommand{\bibinfo}[2]{#2}
\providecommand{\BIBentrySTDinterwordspacing}{\spaceskip=0pt\relax}
\providecommand{\BIBentryALTinterwordstretchfactor}{4}
\providecommand{\BIBentryALTinterwordspacing}{\spaceskip=\fontdimen2\font plus
\BIBentryALTinterwordstretchfactor\fontdimen3\font minus \fontdimen4\font\relax}
\providecommand{\BIBforeignlanguage}[2]{{%
\expandafter\ifx\csname l@#1\endcsname\relax
\typeout{** WARNING: IEEEtran.bst: No hyphenation pattern has been}%
\typeout{** loaded for the language `#1'. Using the pattern for}%
\typeout{** the default language instead.}%
\else
\language=\csname l@#1\endcsname
\fi
#2}}
\providecommand{\BIBdecl}{\relax}
\BIBdecl

\bibitem{ref:Foreign_experts}
H.~Tataria, M.~Shafi, A.~F. Molisch \emph{et~al.}, ``{6G wireless systems: Vision, requirements, challenges, insights, and opportunities},'' \emph{Proc. IEEE}, vol. 109, no.~7, pp. 1166--1199, Jul. 2021.

\bibitem{REF1}
G.~Liu, R.~Xi, Z.~Han \emph{et~al.}, ``{Cooperative sensing for 6G mobile cellular networks: feasibility, performance and field trial},'' \emph{IEEE J. Sel. Areas Commun.}, Oct. 2024.

\bibitem{REF5}
F.~Liu, Y.~Cui, C.~Masouros \emph{et~al.}, ``{Integrated sensing and communications: Toward dual-functional wireless networks for 6G and beyond},'' \emph{IEEE J. Sel. Areas Commun.}, vol.~40, no.~6, pp. 1728--1767, Jun. 2022.

\bibitem{ref:fanTIT}
Y.~Xiong, F.~Liu, Y.~Cui \emph{et~al.}, ``{On the fundamental tradeoff of integrated sensing and communications under Gaussian channels},'' \emph{IEEE Trans. Inf. Theory}, vol.~69, no.~9, pp. 5723--5751, Sep. 2023.

\bibitem{R1_IMT}
I.~R. Sector, ``{Recommendation ITU-R M. 2160-0 Framework and overall objectives of the future development of IMT for 2030 and beyond},'' ITU, Tech. Rep, Tech. Rep., 2023.

\bibitem{REF2}
A.~Liu, Z.~Huang, M.~Li \emph{et~al.}, ``{A survey on fundamental limits of integrated sensing and communication},'' \emph{IEEE Commun. Surv. Tutorials}, vol.~24, no.~2, pp. 994--1034, Feb. 2022.

\bibitem{ref:jialincommag}
J.~Zhang, J.~Wang, Y.~Zhang \emph{et~al.}, ``{Integrated sensing and communication channel: Measurements, characteristics, and modeling},'' \emph{IEEE Commun. Mag.}, vol.~62, no.~6, pp. 98--104, Jun. 2023.

\bibitem{ref:3DMIMOJSAC}
J.~Zhang, Y.~Zhang, Y.~Yu \emph{et~al.}, ``{3-D MIMO: How much does it meet our expectations observed from channel measurements?}'' \emph{IEEE J. Sel. Areas Commun.}, vol.~35, no.~8, pp. 1887--1903, Jun. 2017.

\bibitem{REF2_1_1}
J.~Zhang, J.~Lin, P.~Tang \emph{et~al.}, ``{Channel measurement, modeling, and simulation for 6G: A survey and tutorial},'' \emph{arXiv preprint arXiv:2305.16616}, 2023.

\bibitem{ref:liufanJSAC}
F.~Dong, F.~Liu, S.~Lu \emph{et~al.}, ``{Communication-assisted sensing in 6G networks},'' \emph{IEEE J. Sel. Areas Commun.}, 2025.

\bibitem{REF_2_4}
N.~N. Youssef, ``{Radar cross section of complex targets},'' \emph{Proc. IEEE}, vol.~77, no.~5, pp. 722--734, May 2002.

\bibitem{ref:wenjunletter}
W.~Chen, Y.~Zhang, Y.~Liu \emph{et~al.}, ``{An empirical study of ISAC channel characteristics with human target impact at 105 GHz},'' \emph{Electron. Lett.}, vol.~60, no.~17, p. e70017, Sep. 2024.

\bibitem{REF_4}
\BIBentryALTinterwordspacing
3GPP, ``{Study on Integrated Sensing and Communication (ISAC) in 5G and Beyond},'' {3rd Generation Partnership Project (3GPP)}, Technical Report TR 38.886, Jun. 2021. [Online]. Available: \url{https://www.3gpp.org/}
\BIBentrySTDinterwordspacing

\bibitem{3GPPPPPP}
\BIBentryALTinterwordspacing
------, ``Study on channel model for frequencies from 0.5 to 100 ghz,'' 3rd Generation Partnership Project (3GPP), Technical Report TR 38.901, Mar. 2022. [Online]. Available: \url{https://www.3gpp.org/}
\BIBentrySTDinterwordspacing

\bibitem{ref:yamengISACmodel}
Y.~Liu, J.~Zhang, Y.~Zhang \emph{et~al.}, ``{How to extend 3D GBSM to integrated sensing and communication channel with sharing feature?}'' \emph{IEEE Wireless Commun. Lett.}, Aug. 2024.

\bibitem{ref:yamengshare}
------, ``{A shared cluster-based stochastic channel model for integrated sensing and communication systems},'' \emph{IEEE Trans. Veh. Technol.}, vol.~73, no.~5, pp. 6032--6044, May 2023.

\bibitem{REF_dianci1}
G.~Cakir, M.~Cakir, and L.~Sevgi, ``{An FDTD-based parallel virtual tool for RCS calculations of complex targets},'' \emph{IEEE Antennas Propag. Mag.}, vol.~56, no.~5, pp. 74--90, Dec. 2014.

\bibitem{REF_dianci2}
E.~Lucente, A.~Monorchio, and R.~Mittra, ``{An iteration-free MoM approach based on excitation independent characteristic basis functions for solving large multiscale electromagnetic scattering problems},'' \emph{IEEE Trans. Antennas Propag.}, vol.~56, no.~4, pp. 999--1007, Apr. 2008.

\bibitem{REF_dianci3}
E.~A. Dunn, J.-K. Byun, E.~D. Branch \emph{et~al.}, ``{Numerical simulation of BOR scattering and radiation using a higher order FEM},'' \emph{IEEE Trans. Antennas Propag.}, vol.~54, no.~3, pp. 945--952, Mar. 2006.

\bibitem{REF_dianci4}
F.~Weinmann, ``{Ray tracing with PO/PTD for RCS modeling of large complex objects},'' \emph{IEEE Trans. Antennas Propag.}, vol.~54, no.~6, pp. 1797--1806, Jun. 2006.

\bibitem{REF_dianci5}
T.-Q. Fan and L.-X. Guo, ``{OpenGL-based hybrid GO/PO computation for RCS of electrically large complex objects},'' \emph{IEEE Antennas Wirel. Propag. Lett.}, vol.~13, pp. 666--669, Apr. 2014.

\bibitem{REF_dianci6}
P.~Y. Ufimtsev, ``{Improved physical theory of diffraction: Removal of the grazing singularity},'' \emph{IEEE Trans. Antennas Propag.}, vol.~54, no.~10, pp. 2698--2702, Oct. 2006.

\bibitem{REF12}
W.~Tang, Z.-W. Xu, H.-S. Zhao \emph{et~al.}, ``{Scattering of UAV Swarm via Artificial Plasma Cloud for Probable Over-the-Horizon Detection at VHF Band},'' \emph{IEEE Trans. Antennas Propag.}, May 2024.

\bibitem{REF12_again}
Z.~Yuan, L.~Yu, Z.~Wang \emph{et~al.}, ``{Experimental Analysis and Modeling of Mono-static UAV RCS for ISAC Channels},'' \emph{IEEE Antennas Wirel. Propag. Lett.}, Nov. 2024.

\bibitem{REF13}
M.~Rosamilia, A.~Balleri, A.~De~Maio \emph{et~al.}, ``{Radar detection performance prediction using measured UAVs RCS data},'' \emph{IEEE Trans. Aerosp. Electron. Syst.}, vol.~59, no.~4, pp. 3550--3565, Aug. 2022.

\bibitem{REF15}
T.~Motomura, K.~Uchiyama, and A.~Kajiwara, ``{Measurement results of vehicular RCS characteristics for 79GHz millimeter band},'' in \emph{2018 IEEE Topical Conference on Wireless Sensors and Sensor Networks (WiSNet)}.\hskip 1em plus 0.5em minus 0.4em\relax IEEE, Jan. 2018, pp. 103--106.

\bibitem{REF16}
S.~Abadpour, M.~Pauli, C.~Schyr \emph{et~al.}, ``{Angular resolved rcs and doppler analysis of human body parts in motion},'' \emph{IEEE Trans. Microwave Theory Tech.}, vol.~71, no.~4, pp. 1761--1771, Apr. 2022.

\bibitem{REF17}
J.~Li, D.~Zhang, Z.~Wu \emph{et~al.}, ``{SBRF: A fine-grained radar signal generator for human sensing},'' \emph{IEEE Trans. Mob. Comput.}, vol.~23, no.~12, pp. 13\,114--13\,130, Dec. 2024.

\bibitem{REF18}
Y.-X. Zhang, Y.-C. Jiao, M.-D. Zhu \emph{et~al.}, ``{Wideband near-field RCS measurement techniques with improved far-field RCS prediction accuracies},'' \emph{IEEE Trans. Instrum. Meas.}, vol.~72, pp. 1--4, Dec. 2022.

\bibitem{REF19}
T.~Schipper, J.~Fortuny-Guasch, D.~Tarchi \emph{et~al.}, ``{RCS measurement results for automotive related objects at 23-27 GHz},'' in \emph{Proceedings of the 5th European Conference on Antennas and Propagation (EUCAP)}.\hskip 1em plus 0.5em minus 0.4em\relax IEEE, Apr. 2011, pp. 683--686.

\bibitem{ref:116meeting}
\BIBentryALTinterwordspacing
3GPP, ``Chair notes, {3GPP TSG RAN WG1 \#116}, meeting report {TSGR1\#116},'' 3rd Generation Partnership Project (3GPP), Meeting Report TSGR1\#116, Feb. 2024. [Online]. Available: \url{https://www.3gpp.org/ftp/tsg_ran/WG1_RL1/TSGR1_116/Inbox/Chair_notes}
\BIBentrySTDinterwordspacing

\bibitem{ref:peinews}
Y.~Zhang, J.~Zhang, Y.~Pei \emph{et~al.}, ``Latest progress for {3GPP} {ISAC} channel modeling standardization,'' \emph{Science China: Information Sciences}, vol.~67, no.~11, pp. 357--358, Oct. 2024.

\bibitem{ref:ISACScenarios}
\BIBentryALTinterwordspacing
{AT\&T}, ``{R1-2410337}: {FL} summary \#1 on {ISAC} deployment scenarios,'' 3rd Generation Partnership Project (3GPP), Orlando, US, Working Group Document TSGR1\#119, Nov. 2024. [Online]. Available: \url{https://www.3gpp.org/ftp/tsg_ran/WG1_RL1/TSGR1_119/Docs/R1-2410337.zip}
\BIBentrySTDinterwordspacing

\bibitem{REF11}
S.~A. Hovanessian, ``{Radar system design and analysis},'' \emph{Dedham}, 1984.

\bibitem{REF_duomian}
W.~B. Gordon, ``{Near field calculations with far field formulas},'' in \emph{IEEE Antennas and Propagation Society International Symposium. 1996 Digest}, vol.~2.\hskip 1em plus 0.5em minus 0.4em\relax IEEE, Jul. 1996, pp. 950--953.

\bibitem{REF211_KL}
J.~Jiao, T.~A. Courtade, A.~No \emph{et~al.}, ``{Information measures: the curious case of the binary alphabet},'' \emph{IEEE Trans. Inf. Theory}, vol.~60, no.~12, pp. 7616--7626, Dec. 2014.

\bibitem{REF21}
\BIBentryALTinterwordspacing
3GPP, ``{ISAC} channel measurements and modeling,'' 3rd Generation Partnership Project (3GPP), Working Group Document R1-2406107, Aug. 2024, 3GPP TSG RAN WG1 Meeting \#118, Orlando, US, August 19th--23rd, 2024. [Online]. Available: \url{https://www.3gpp.org/ftp/TSG_RAN/WG1_RL1/TSGR1_118/Docs/R1-2406107.zip}
\BIBentrySTDinterwordspacing

\bibitem{REF_pingtai}
\BIBentryALTinterwordspacing
{BUPT-CMCC}, ``{BUPTCMCCCMG-IMT2030} channel model platform,'' Online Platform, Feb. 2023. [Online]. Available: \url{https://scc.bupt.edu.cn/dataset-public/datasets/22}
\BIBentrySTDinterwordspacing

\bibitem{REF_ITU}
\BIBentryALTinterwordspacing
{ITU-R}, ``Guidelines for evaluation of radio interface technologies for {IMT-2020},'' International Telecommunication Union - Radiocommunication Sector (ITU-R), Recommendation M.2412, Oct. 2017. [Online]. Available: \url{https://www.itu.int/}
\BIBentrySTDinterwordspacing

\end{thebibliography}

\end{document}